\documentclass[12pt]{extarticle} 

\usepackage [T1]{fontenc}
\usepackage {calligra}
\usepackage [T1]{pbsi}
\usepackage{CJKutf8} 
\usepackage[colorlinks=true,linkcolor=blue]{hyperref}
\usepackage{amsmath,amssymb,amsfonts} 
\usepackage{graphicx}                 
\usepackage{color}                    
\usepackage{hyperref}                 
\usepackage[left=1cm,top=1cm,bottom=1cm,right=1cm,nohead,nofoot]{geometry} 
\definecolor{red}{rgb}{1,0,0}           
\definecolor{green}{rgb}{0,1,0}
\definecolor{blue}{rgb}{0,0,1}
\definecolor{darkblue}{rgb}{0,0,0.5}
\definecolor{lightblue}{rgb}{.5,.5,1}
\definecolor{lightgray}{gray}{.87}          
\definecolor{Dark}{gray}{.20}
\definecolor{pink}{rgb}{.95,0.82,0.92}  
\definecolor{yellow}{rgb}{1,1,0}
\definecolor{lightyellow}{rgb}{1,1,.5}
\definecolor{purple}{rgb}{0.7,0,0.85}
\definecolor{darkgreen}{rgb}{0,0.5,0}
\definecolor{orange}{rgb}{0.8,0.2,0.2}
\def \be {\begin{equation}}
\def \ee {\end{equation}}
\def \bea {\begin{eqnarray}}
\def \eea {\end{eqnarray}}
\def \nn {\nonumber}

\def \rr {\raise.35ex\hbox{\small $\prime$}\kern-.17em{\mbox{\large $\imath$}}}
\def \del {\partial}
\def \dels {\partial\kern-.5em / \kern.5em}
\def \As {{A\kern-.5em / \kern.5em}}
\def \Ds {D\kern-.7em / \kern.5em}

\def \a {\alpha}

\def \b {\beta}

\def \g {\gamma}

\def \d {\delta}
\def \eps {\epsilon}
\def \m {\mu}
\def \n {\nu}
\def \k {\kappa}
\def \lam {\lambda}
\def \Lam {\Lambda}

\def \II {I\hspace{-.1em}I\hspace{.1em}}

\def \IIA {\mbox{\II A\hspace{.2em}}}

\def \z {(0)}
\def\nz{(KK)}

\newcommand{\solution}[1]{}

\begin{document}


\fontsize{12pt}{13pt}\selectfont

\pagestyle{plain}

\begin{CJK}{UTF8}{bsmi}






\begin{titlepage}

\begin{center}
\hfill.
\vskip .2in

\textbf{\LARGE Non-Abelian Chiral 2-Form and M5-Branes
}

\vskip .5in
{\large
Kuo-Wei Huang\footnote{
E-mail: kuo-wei.huang@stonybrook.edu }
}\\
\vskip 2mm
{\it\large
Department of Physics and Center for Theoretical Sciences,\nn\\ National Taiwan University, Taipei 10617, Taiwan}\\

\vspace{75pt}
\end{center}
\begin{abstract}

We first review self-dual (chiral) gauge field theories by studying their Lorentz non-covariant and
Lorentz covariant formulations. We next construct a non-Abelian self-dual two-form gauge theory
in six dimensions with a spatial direction compactified on a circle.  This
model reduces to the Yang-Mills theory in five dimensions for a small compactified radius R. The
model also reduces to the Lorentz-invariant Abelian self-dual two-form theory when the
gauge group is Abelian. The model is expected to describe multiple 5-branes in M-theory. We also discuss its decompactified limit, covariant formulation, BRST-antifield quantization and other generalizations.

\end{abstract}

\end{titlepage}

\newpage
\tableofcontents

\newpage
\section{Introduction}
A self-dual gauge field theory (or a chiral gauge field theory) is defined as a $p$-form gauge potential whose $p+1$ form field strength is constrained by the self-duality condition, which reduces physical degrees of freedom in a theory to the half of the case without the self-duality condition. If we consider our spacetime with the Lorentz signature, real self-dual gauge fields can exist only in 4n+2 dimensions like in D=2, D=6 or D=10, which are the cases we will  consider in this thesis. Self-dual gauge field theories have received huge attention for quite a long time because of their existence in various interesting theories. For examples, they appear in the quantum hall effect in solid state physics and also show up in the heterotic string theory. They are fundamental elements in the exotic six-dimensional 5-brane(s) worldvolume theory in M-theory (or in NS 5-brane(s) worldvolume theory in the type IIA string theory). Self-dual fields also appear in the ten dimensional type IIB supergravity. However, formulating a theory for self-dual gauge fields is challenging because  a standard action does not really work. We will discuss this issue in detail. Self-dual gauge fields also have subtle issues regarding their quantization. But we will largely focus on the classical level.  In fact , we are going to face even bigger challenges when trying to describe multiple M5-branes worldvolume theory, which is our main topic.  (This thesis is based on \cite{HHM}. We will provide more detailed discussion and some related  generalizations.).
\\

Our motivation comes from string/M theory. Five string theories in ten dimensions can be unified by the eleven dimensional M-theory, where the basic objects are M2-branes and M5-branes (which can be viewed as the magnetic dual of M2-branes in the sence that M5-branes couple to the dual background three form C-field in 11D supergravity). Strings and D-branes can be viewed as compactified M-branes. Therefore, one could say that M-branes are the most fundamental (theoretical) objects we have. (For some review papers of M-theory branes, see \cite{review}.)  In order to understand the mysterious nature of the M-theory, it is certainly desirable to better understand the bacis properties (symmetry structure) of these M-branes. The description of a single M2-brane or a single M5-brane had already known for quite a long time. The understanding of multiple M2 branes gained ground in the past few years by the so-called BLG and ABJM theory \cite{BLG} \cite{ABJM}. It is reasonable to start thinking about the formulation of multiple M5-branes.
\\

The physical degrees of freedom of the 6-dimensional worldvolume theory of a M5-brane \cite{mM5} consist of the so-called $N = (2, 0)$ tensor supermultiplet, which contains two-form gauge potentials with self-dual field strengths, five scalars, and two chiral spinors. These scalars and spinors can be interpreted as Goldstone bosons and fermions associated with broken translation symmetries and supersymmetries. When the 11-dimensional M theory is compactified on a circle, it effectively reduce to the 10-dimensional type IIA superstring theory if the radius of the circle $R\to 0$. D-branes in the IIA theory have M-theory interpretations. In particular, wrapping one dimension of M5-branes on the compact spatial dimension gives the IIA D4-branes theory.  This is one of the main criteria to construct M5-branes theory. Throughout this thesis, we will focus on the gauge field sector. We also believe that the gauge field's part brings the most subtle issues.
\\

When several D-branes start to coincide, the Maxwell field (U(1) gauge theory) living in a D-brane is enhanced and become Yang-Mills gauge fields. In short, multiple D-branes in string theory can be effectively decribed by a non-abelian gauge theory. Therefore, one would expect that some kind of non-abelian 2-form gauge theory will be involved in the multiple M5-branes system. However, we expect that M5-branes can not be a simple non-Abelian gauge theory because it is found that the entropy (degrees of freedom) of N coincident M5-branes does not scale as $N^2$ like in a standard Yang-Mills theory.  The entropy of M5-branes scales as $N^3$ \cite{N3}. \footnote{Also see a recent derivation \cite{Maxfield:2012aw} of the D=6 conformal anomaly which also provides a way to compute the $N^3$ of M5-branes.} In the case of the theory of multiple M2-branes, it is also not an ordinary non-Abelian theory (the entropy of N coincident M2-branes scales as $N^ {3\over2}$ ) \cite{BLG} \cite{ABJM}. It is natural to ask whether or not the similar structure neeed for M2-branes (for example, Lie 3-algebra used in \cite{BLG}) could play also a crucial role in multiple M5-branes. However, although there are some approaches along this line (See, for example, \cite{m5 other},\cite{Kawamoto:2011ab},\cite{Honma:2011br}.), it is not clear how to relate to the hidden non-Abelian gauge symmetry structure of multiple M5-branes. 
\\

Some believed that a Lagrangian formulation for self-dual gauge theories
was impossible, mainly because the self-duality condition imposes first order differential equations
on the gauge potentials, while an ordinary kinetic term in a standard Lagrangian (take 2-form potentials in 6D as an example)
\bea
L\sim H_{\m\n\lam}H^{\m\n\lam}\sim \del_{[\mu} B_{\nu\lam]}\del^{\mu}B^{\nu\lam} \ ,
\eea
(where $H_{\m\n\lam}$ is the field strength of 2-form potentials and $\m,\n= 0,1,2,3,4,5$) always leads to a 2nd order differential equation.
It turns out that the trick is to avoid using some of components
of the gauge potentials, for example, components $B_{i5}$ (where $i,j = 0,1,2,3,4$) will enter the action only up to surface terms. So even though we get 2nd order differential equations from varying the action, the self-duality condition appears after integrating once the equations of motion. This trick was later generalized in \cite{Chen:2010jgb}
so that for a given spacetime dimension $D$, one can write down a Lagrangian for self-dual gauge fields for arbitrary divisions of $D$ into two positive integers $D'$ and $D''$ \bea D' + D'' = D \ .\eea
We call it as the $(D' + D'')$-formulation of
self-dual gauge theories. (Some non-covariant Lagrangians based on decompositions of spacetime $D=D_1+D_2+D_3$ are later considered in \cite{WHH}.) Hence the Lagrangian for a self-daul gauge field theory was first constructed
$without$ the manifest Lorentz symmetry. Lorentz-covariant formulations are possible only when one
introduces auxiliary fields \cite{PST,oldM5}. 
\\

The gauge symmetry for a single M5-brane in the trivial background is Abelian.
The first non-Abelian gauge theory for self-dual 2-form potentials
was found when considering an M5-brane in a large $C$-field background \cite{NPM5}. We call it as $NP$ M5-brane theory, where
''$NP$'' stands for ''Nambu-Poisson''. For a brief review about this topic, see \cite{HoM5}. A Nambu-Poisson structure is used to define the non-Abelian gauge symmetry
for the 2-form potential on a M5-brane. The physical origin of this Nambu-Structure is the coupling of open membranes
to the $C$-field background \cite{ToyModel}.  The NP M5-brane theory was first derived from the BLG model \cite{BLG}. Its gauge field content was further explored in \cite{Pasti,Furuuchi}.  A double dimension reduction of the NP M5-brane theory is in agreement with the lowest order deformation of
the noncommutative D4-brane action in large NS-NS $B$ field background \cite{NPM5}.
Thus if this NP M5-brane theory can be properly deformed such that it agrees with the noncommutative D4-brane theory to ''all orders'',
it might resemble the non-abelian structure of multiple M5-branes theory, since the multiple D4-branes theory is essentially a special case of the noncommutative D4-brane theory. However, it turns out that it is extremely hard
to deform the NP M5-brane theory \cite{Chen:2010ny}.  We conclude that it takes brand new ideas to construct multiple M5-branes theory.
\\

In the literature, there has been various attempts to construct a non-Abelian gauge theory for 2-form gauge potentials $B_{\m\n}$.
Taking values in a Lie algebra, the corresponding geometrical structures are called ''non-Abelian gerbes''. However, the immediate problem to construct such a model is that one needs to define covariant derivatives $D_{\mu}$ to have gauge covaraint structures in a theory,
thus we also need a 1-form potential $A$.
For example, in \cite{AB}, the gauge transformations of $A$ and $B$ are defined by
\bea
A' &=& gAg^{-1}+gdg^{-1} + \Lambda, \label{nag1}\\
B' &=& gBg^{-1}+\left[ A',\Lambda\right]_{\wedge} + d\Lambda+\Lambda\wedge\Lambda,
\label{nag2}
\eea
where $g\in G$ is the gauge parameter and $\Lambda\in g$ is a 1-form.
Mathematically such gauge transformations are well-defined and suitable to describe some systems such as the non-Abelian
generalization of the BF model \cite{BF}. It is, however, not clear if it is relevant to describe multiple M5-branes.
\\

Physically, the introduction of an extra field like $A$ increases
the physical degrees of freedom of the system.
For the M5-branes system,
there is no physical degrees of freedom corresponding to any new field.
Furthermore, with the addition of $A$,
the field $B$ is not a genuine 2-form potential in the sense that
we can gauge away $A$ by $\Lambda$,
and then $B$ is not independent of its longitudinal components.
The result is similar to spontaneous symmetry breaking. On the other hand, if we consider that A does not have physical degrees of freedom, it means that we might have a gauge symmetry to gauge it away. But this means that the covariant derivative in a theory become just $\del_{\mu}$. It will be problematic to define the non-abelian gauge transformation and the field strength  in this gauge. (One attempt to construct non-Abelian 2-form gauge theory
is to define it on the loop space \cite{loopspace}.
This approach introduces infinitely many more degrees of freedom in a theory.)
\\

Our goal here is to have a non-Abelian 2-form gauge symmetry
which includes the Abelian 2-form theory as the special case
when the Lie algebra involved is Abelian.
This criterium is not matched by any existing construction
in literatures. Note that assuming the existence of an action and gauge transformation laws, a no-go theorem \cite{no-go}
states that it does not exist any nontrivial deformation of
the Abelian 2-form gauge theory. We notice that one of their assumptions was  the "{\em locality}'' for the action and the gauge transformation laws. Also, the Lorentz symmetry was {\em not} assumed.
\\

The non-existence of a local action for multiple M5-branes
was argued in another way by Witten \cite{Witten}.
The M5-branes system is known to have conformal symmetry,
which implies that upon double dimension reduction,
the 4+1 dimensional action should be proportional to
$ \int d^5 x {1\over R}. \label{W} $
On the other hand,
the reduction of a 5+1 dimensional local action on a circle should give
$ \int d^5 x ~2\pi R $, 
which has the opposite dependence on $R$.
As long as we assume a Lorentz-covariant formulation for M5-branes
without explicit reference to the compactifiation radius $R$
except through the measure of integration, this gives a strong argument against
the Lagrangian formulation of multiple M5-branes.
\\

Recently, there are proposals \cite{Douglas,LPS} claiming that
the multiple M5-branes compactified on a circle of finite radius $R$
could be described by the $U(N)$ super Yang-Mills (SYM) theory for $N$ D4-branes
even before taking the small $R$ limit. This would be a duality between two theories in 5 and 6 dimensions, respectively,
but it can not be viewed as an example of the holographic principle of quantum gravity,
because there is no gravitational force in these theories.
However, we will point out its difficulties later.
\\

The above mentioned developments suggest that
it is already a tremendous progress to try to have a theory for multiple M5-branes compactified on a circle of finite radius $R$. We stress that the following two criteria must be satisfied:
\begin{enumerate}
\item
In the limit $R\rightarrow 0$,
the theory should be approximated by the gauge field sector of 
the multiple D4-branes theory,
which is the $U(N)$ Yang-Mills theory in 5 dimensions.
\item
When the Lie algebra of the gauge symmetry is Abelian,
the theory should reduce to $N$ copies of the Abelian self-dual 2-form gauge field theory. (Single M5-brane theory.)
\end{enumerate}

In view of no-go theorems \cite{no-go},
the absence of 6 dimensional Lorentz symmetry
due to compactification of the 5-th direction
does not really necessarily make the task easier.
(Notice that the 2nd criterion ensures the 6 dimensional Lorentz symmetry
in the broken phase in the limit $R \rightarrow \infty$.)
In the following we will construct an interacting theory
satisfying both criteria. The cost we have to pay to meet these criteria is a nonlocal treatment 
of the compactified dimension as we will see in the following sections.
Such a description may seem exotic, 
but it might be justified 
in view of the special role played by the compactified direction
in defining M-theory as the strong coupling limit of type $\IIA$ string theory.
\\

We also notice that there are some recent works on non-Abelian two-form theory and M5-branes \cite{more M5}, \cite{Chu:2012um}, but we will not be able to review them in this thesis \footnote{The model given in \cite{Chu:2012um} can be viewed as a gauge-fixed version of \cite{HHM}.}. On the other hand, for a more general approach to self-dual field theories and has the advantage to deal with subtle topological issues, we refer to the work of Belov and Moore \cite{moore}. It will be interesting to relate the model in this thesis to their formulation.
\\

We organize this thesis as follows.
\\

In section 2, we start from the simplest example of a self-daul field theory:  The chiral boson in two-dimensions. We will argue that why some some naive methods to construct a self-dual theory do not really work. We then discuss two approaches to obtain a chiral boson action. Here we will  observe  some universal properties of self-dual theories such as extra gauge symmetries and modified Lorentz transformation laws. 
\\

In section 3, we first introduce the standard 2-form potential gauge theory. We briefly discuss the BF theory in four dimensions to familiarize ourself with some basic structure of 2-form gauge fields. We then start to consider the formulation of the self-dual 2-form theory for a single M5-brane. The model we follow here is referred as PS (Perry and Schwarz) formulation \cite{PeS}.\footnote{I thank M. Henneaux for reminding me that the action considered in this section was written long ago \cite{HT} for the first time. Its Lorentz invariance was established there also.} We will also study different (Abelian) self-dual 2-form theories under different spacetime decompositions. 
\\

In section 4, we study Lorentz-covaraint versions of the 2D chiral boson and the 6D chiral 2-form with the help of introducing auxiliary fields. These covariant formulations are given by \cite{PST} and will be referred as PST (Pasti, Sorokin and Tonin) formulations. We will see how these covaraint actions consistently reduce to the previous non-covariant actions via gauge fixing conditions. 
\\

In section 5, which is the body of this thesis, we formulate a non-abelian 2-form gauge theory. We non-abelianize the PS action and show how this non-Abelian action gives the self-dual equation. In particular, we will also see this model indeed reduces to the Yang-Mills theory in 5 dimensions for small compactified radius R and it reduces to the Lorentz-invariant theory of Abelian chiral 2-forms theory when the gauge group is Abelian. 
\\

In section 6, the BRST transformation laws of this non-abelian 2-form gauge algebra are given. A crucial point is that we will need to introduce the ghost of the ghost for this kind of reducible gauge system. A BRST invaraint gauge-fixed action is also obtained by utilizing the BRST-antifield (BV) method.
\\

In section 7, we generalize our method to construct a non-Abelian gauge theory for 3-form gauge potentials.
\\

We conclude this thesis in section 8 by pointing out difficulties when trying to find a manifest PST-like Lorentz-covariant formulation of the non-abelian self-dual 2-form theory if using the similar idea of introducing auxiliary fields. We will also  point out some main open problems.
 
\newpage
\section{A Toy Model: 2D Chiral Boson}

The simplest example of a self-daul field theory is in two-dimensions: 2D chiral bosons. They play an essential ingredient in the quantum Hall effect or the construction of heterotic-type string theory which is phenomenologically interesting because of the left-right asymmetry.
\\

In 2D, the anti-symmetric tensor is just a scalar $\phi$ and its field strength is defined by
\bea
F_a=\del_a \phi \ ~~,~~ a=0,1 \ .
\eea It is obvious that there is a global symmetry 
\bea \delta \phi= \mbox {constant} \ .
\eea 
The self-duality condition is defined by
\bea
\del_0 \phi-\del_1\phi\equiv {\cal F}=0 \ .
\eea  The first quesiton is: Can we have an action that gives this self-duality condition? First of all, let us learn something from making mistakes. If one wants to write down an action that gives the above self-duality constriant equation, a nature approach might be to introduce a Lagrange-multiplier field $\lam$ to implement this constraint using the following action 
\bea
S=\int d^2x \Big[-{1\over2}F_a F^a+\lam{\cal F}\Big] \ .
\eea 
Indeed, the self-duality condition is obained via the variation of the action with respect to $\lam$. However, this method is not correct since the Lagrange multiplier term itself ends up generating an extra field equation from the field equation of the scalar $\phi$.
\\

Here is a clever trick, introduced by Siegel (1984) \cite{Siegel}, who suggested to consider a ''squared self-duality contraint'' in an action
\bea
\label{Siegel}
S=\int d^2x \Big[-{1\over2}F_a F^a+\lam({\cal F})^2\Big] \label{1S} \ .
\eea In this case, the field equation of Lagrange-multiplier field $\lam$ still gives the self-duality condition, $\del_0 \phi-\del_1\phi=0$. But the key difference here is that the field equation of $\phi$, which can could be written as
\bea
(\del_0+\del_1){\cal F}-2\lam(\del_0-\del_1){\cal F}=0~~ (\equiv0~~ if~ {\cal F}=0) \ ,
\eea does not give any extra constraint by using the self-duality condition. That is, it vanishes identically hence it gives simply a redundant equation of motion. Furthermore, the redundant field equation occurs us the existence of an additional gauge symmetry in the theory. One could indeed find the following  new gauge transformation that leaves the action \eqref{Siegel} invariant
\bea
\delta\phi&=&\cal F \eps \ , \\
\delta \lam&=&{1 \over 2}(\del_0+\del_1) \eps+\eps(\del_0-\del_1)\lam+\lam (\del_0-\del_1)\eps \ ,
\eea where $\eps$ is the gauge parameter. 
\\

However, although this model shows that the Lagrange multiplier does not cause further constraints, people found Sigel's model, in particular in higher dimensions that we are interested later, suffers from the anomaly problems if not properly treated when quantized \cite{QSD}. 
Furthermore, it is not totally clear if there are enough local gauge symmetries in Seigel-like formulations in $D=6$ case or $D=10$ case in order to completely gauge away the Lagrange-multiplier fields \cite{QSD}. Since the case $D=6$ is of our the main topic, in the following we would like to introduce yet another formulation that could be generalized to higher dimensions without too much trouble. We will see the self-duality condition could be obtained by another interesting way rather than the squared constraint in the Seigel's formulation.
\\

The formulation of (2D) chiral boson we consider was invented by R. Floreanini and R. Jackiw (1987) \cite{FJ}. The FJ action reads
\bea
\label{FJJ}
S_{1+1}=-{1\over 2}\int d^2x ~\del_x \phi (\del_t \phi-\del_x\phi) \ ,
\eea where $1+1$ means that the decomposition of 2D Minkowski space
\bea
R^{1+1}=R^1\times R^1 \ ,
\eea 
which is obviously the only possible decomposition in 2D. The salient property of this action is the existence of the following $gauge$ transformation: 
\bea
\label{gg}
\delta \phi=f(t) \ ,
\eea 
for an arbitrary function $f(t)$ independent of $x$. One can check now:
\bea
\delta S\sim \int d^2x ~\del_x (\phi \del_t f(t))=tot \ .
\eea
So the action is indeed invariant up to a total derivative. (We assume that a surface term can be dropped in our manifold, under some suitable boundary conditions.)  Notice that this symmetry is different from the standard constant  (trivial) transformation. 
From the equation of motion
\bea
\del_x(\del_x \phi-\del_t\phi)=0 \ ,
\eea 
we find it implies
\bea
(\del_x \phi-\del_t\phi)=g(t) \ ,
\eea for an arbitrary function $g(t)$ independent of $x$. So the main trick is that we now can use the new gauge transformation \eqref{gg} by setting
\bea
f(t)=-\int^t dy g(y) \ ,
\eea 
to absorb the function $g(t)$. We therefore arrive the self-duality condition
\bea
\del_x \phi-\del_t\phi=0 \ .
\eea

On other other hand, one notices that the FJ action \eqref{FJJ} is not manifestly Lorentz invariant. But an additional interesting property of this formulation is that the action is invariant under the following modified Lorentz transformation law
\bea
\delta \phi= \lam^{tx}(t\del_x-x\del_t)\phi+\lam^{tx} x(\del_t \phi-\del_x\phi) \ ,
\eea 
where $\lam^{tx}$ is the Lorentz parameter. The standard transformation law is modified by the self-duality condition.

\newpage
\section{Abelian Chiral 2-Form: Single M5-Brane}

\subsection{Anti-Symmetry 2-rank Gauge Field and BF Model}
Let us first introduce our notation through a short review on the anti-symmetry 2-rank gauge fields. We consider $R^{1,5}$ as the 6D Minkowski space parametrized by $x^\m$ with $\m=0,1,2,3,4,5$. The 2-form potential is denoted by $B_{\m\n}$. Antisymmetry property implies that there is no diagonal components in $B_{\m\n}$. The corresponding field strength is defined by
\bea
H_{\m\n\lam}=\del_\m B_{\n\lam}+\del_\n B_{\lam\m}+\del_\lam B_{\m\n} \ ,
\eea 
which is invariant under the following gauge symmetry transformation:
\bea
\delta B_{\m\n}=\del_\m \Lam_\n-\del_\n \Lam_\m \ ,
\eea 
in terms of the 1-form gauge parameter $\Lam_\m$.  Naively, it seems that $\Lam_\m$ has six independent degrees of freedom. However, a crucial property of this (higher rank) gauge symmetry is its redundancy: The gauge transformation itself has a gauge transformation given by
\bea
\delta \Lam_\m=\del_\m \a \ ,
\eea 
parametrized by the 0-form gauge parameter $\a$. So in fact only five of six gauge parameters are independent. 
\\

The action for the standard 2-form gauge potentials is given by
\bea
S=-{1\over 6}\int d^6x~H^{\m\n\lam}H_{\m\n\lam} \ .
\eea 
The field equation is
\bea
\del_\m H^{\m\n\lam}=0 \ .
\eea 
And there is an obvious (Bianchi) identity
\bea
\eps^{\m\n\lam\g\rho\sigma}\del_\lam H_{\g\rho\sigma}=0 \ .
\eea Notice that in 4D, a 2-form potential theory is dual to a scalar potential theory which has only one physical degree of freedom. That is, in four dimensions we have
\bea
L_{4D}=-{1\over 6} H^{\m\n\lam}H_{\m\n\lam}=-{1\over 6}\times {1\over2} \eps_{\m\n\lam\rho} \eps^{\m\n\lam\sigma}\del^\rho\phi\del_\sigma\phi={1\over2}\del_\rho\phi\del^\rho\phi
\eea 
where $\phi$ represents a real scalar field. In 6D that of our interesting, a 2-form potential has six physical degrees of freedom ($C^{6-2}_2=6$). Knowing how to correctly count the physical degrees of freedom is important since the M5-brane(s) has the supersymmetry structure. The self-duality of 2-form potentials is needed in order to have correct degrees of freedom for M5-brane(s) theories. 
\\

It is also interesting to note that, although $B_{\m\n}$ only increases one index compared with one-form potential $A_\m$. Many things that happen in usual theories in terms of one-form $A_\m$ do not have a straightforward generalization. For example, in the Yang-Mills theory, interaction terms like $\epsilon^{\m\n\lam\rho} [A_\m, A_\n]$ can not have its simple generalization to $\epsilon^{\m\n\lam\rho\sigma\tau} [B_{\m\n}, B_{\lam\rho}]$, because it is zero by itself. Similarly, even in the abelian case where the Chern-Simon term for ${A_\m}$ is given by 
\bea 
L\sim\eps^{\m\n\lam}A_\m \del_\n A_\lam \ ,  
\eea which gives the field equation 
$ F_{\m\n}=0 $ \ ,
a naive generalization to the 2-form potentials
\bea 
L\sim\eps^{\m\n\lam\rho\sigma\tau} B_{\m\n} \del_\lam B_{\sigma\tau} \ ,
\eea  gives a trivial (vanishing) field equation, i.e. 0=0.
\\

A famous model in four-dimensions using two-form potentials $B_{\m\n}$ and is quite different from the standard action is the so-called  BF model. Let us briefly review it. The abelian action is given by
\bea
S= \int d^4x ~\eps^{\m\n\lam\rho} B_{\m\n}F_{\lam\rho} \ ,
\eea 
where $F_{\m\n}$ is Maxwell field strength. Notice that the action does not involve a spacetime metric and one can see that the action is invariant (up to surface terms) under the transformation
\bea
\delta A_\m=\del_\m \lambda~;~
\delta B_{\m\n} = \del_\m \Lambda_\n- \del_\n \Lambda_\m \ .
\eea 
The equation of motions obtained by varying the action with respect to $B_{\m\n}$ and $A_{\m}$ are
$F_{\m\n}=0, H_{\m\n\lam}=0$. The solutions of field equations are equivalent to $A_\m=0=B_{\m\n}$ (assuming the manifold is topological trivial). We see that there is no local degree of freedom in the theory. 
\\

Next we consider a generalization of the above action 
\bea
S= \int d^4x ~\eps^{\m\n\lam\rho} (B_{\m\n}F_{\lam\rho}-{1 \over 2}B_{\m\n}B_{\lam\rho})
\eea 
where the second term is the cosmological term. The action is obvious invariant under $\delta A_\m=\del_\m \lam$. The second term, however, breaks the symmetry $\delta B_{\m\n} = \del_\m \Lambda_\n- \del_\n \Lambda_\m$. But one notices that this action is invariant under a larger gauge symmetry
\bea
\delta A_\m=\omega_\m~;~
\delta B_{\m\n}=\del_\m \omega_\n- \del_\n \omega_\m \ .
\eea 
Thus, one can use this gauge to locally set $ A_\m=0$.
The equation of motions obtained by varying the action with respect to $B_{\m\n}$ and $A_{\m}$ are given by
\bea
F_{\m\n} =B_{\m\n}~;~ H_{\m\n\lam} =0 \ ,
\eea 
where the second equation is simply a consequence of the first one by recalling the Bianchi identity. We see again that there is also no local degrees of freedom in the theory. 
\\

Now we consider the non-abelain generalization. The action is given by
\bea
S= \int d^4x~ Tr~\eps^{\m\n\lam\rho} B_{\m\n}F_{\lam\rho}
\eea 
where $F_{\m\n}=[D_\m,D_\n]$ is the usual Yang-Mills field strength. The action is invariant under the following gauge transformation
\bea
\delta A_\m=0~;~
\delta B_{\m\n}= [D_\m,\Lambda_\n]-[D_\n,\Lambda_\m] \label{gau} \ ,
\eea 
in terms of the 1-form gauge parameter. The action is also invariant under the following gauge transformation
\bea
\delta A_\m=[D_\m,\lambda]~;~
\delta B_{\m\n}= [B_{\m\n},\lambda] \ ,
\eea 
in terms of a 0-form gauge parameter. The action is also invariant under the transformation
\bea
\delta A_\m=0~;~
\delta B_{\m\n}=[F_{\m\n},\omega] \ ,
\eea 
using another 0-form gauge parameter. The equation of motions obtained by varying the action with respect to $B_{\m\n}$ and $A_{\m}$ are given by
\bea
F_{\m\n} = 0~;~ \eps ^{\m\n\lam}[D_\m, B_{\n\lam} ]=0
\eea 
which means field A is a flat connection
On the other hand, the solution of the second equation above can be given by
\bea
B_{\m\n}=\del_\m \Phi_\n- \del_\n \Phi_\m \ ,
\eea 
for some vector $\Phi$. Now we observe that $B_{\m\n}$ can be set as zero by the gauge transformation (\ref{gau}). We see again that there is no local degree of freedom in the theory. 
\\

In sum, BF-like theories are topological field theories and it can be generalized to any dimension. (It is not necessarily that one always need to use two-form potentials in BF theories.) BF theories have some deeper connections to Chern Simons theory, knot theory and also gravitational theories. It is also used to construct mass generation formulation for 1-form gauge potential as an alternative approach of the Higgs mechanism. But we should not further discuss these topics. A motivation to study BF-like theory is to see if M5-branes theory might be related to some topological theories in 6D. However, we will not explore along this direction in this thesis.
\\

Now we turn to our main topic of this thesis: Self-dual 2-form theory.  As mentioned, self-duality means that the original six physical degrees of freedom are reduced to three physical degrees of freedom if imposing the following self-duality condition
\bea
H_{\m\n\lam}=\tilde H_{\m\n\lam}\equiv {1\over 6}\eps_{\m\n\lam\g\rho\sigma}H^{\g\rho\sigma} \ .
\eea 
(One could consider a different sign above as the anti-self duality condition. But it does not really matter which sign we select.)  As we have seen in the 2D chiral boson case, a general structure of chiral field formulation is the existence of an extra gauge symmetry. In the following subsections, we will study 6D (Abelian) self-dual actions under various spacetime splitting decompositions. All these actions can lead to the self-duality condition as the consequence of field equations of 2-form potentials using new gauge symmetries.

\subsection{5+1 Splitting Formulation}

This kind of decomposition is the simplest case in 6D, where only one direction will be treated differently (Here we follow \cite{PeS},\cite{HT}).   In this formulation, two-form potentials live in the following 5+1 decomposition of the spacetime
\bea R^{1,5}=R^{1,4}\times R^1 \ .\eea 
The gauge potentials are decomposed into two sets:
\bea
B_{\m\n}= (B_{i5},~B_{ij}) \ ,
\eea where $i,j= (0,1,2,3,4)$ and $5$-th is the special direction.  Notice that we do $not$ need to consider any compactification at this state. One can also select the time direction as the special one, but our selection will be convenient when it comes to the dimensional reduction to a D4-brane if we need.
\\

Let us first denote the following notations
\bea
{\cal H}_{ij5}&\equiv& H_{ij5}-\tilde H_{ij5}=H_{ij5}+{1\over 6}\eps_{ijnpq}H^{npq} \ , \\
{\cal H}_{ijk}&\equiv& H_{ijk}-\tilde H_{ijk}=H_{ijk}-{1\over 2}\eps_{ijklm}H^{lm5} \ .
\eea 
We also have
\bea
{\cal H}_{ij5}={1\over 6}\eps_{ijnpq}{\cal H}^{npq} \ .
\eea
The action for the abelian chiral two-form in 6D is given by
\bea
\label{51}
S_{5+1}&=&-{1\over12} \int d^6x~\Big[\eps_{ijklm} H^{klm}{\cal H}^{ij5}\Big]\nn\\
&=&{1\over12} \int d^6x~\Big[H_{\m\n\lam}H^{\m\n\lam}-3{\cal H}_{ij5}^2\Big] \ ,
\eea 
where the second expression will be useful when considering the Lorentz-covariant formulation later. But in this subsection we will adopt on the first expression.
\\

We first notice that the manifest Lorentz invariance maintains only in five dimensions. However, just like previous 2D chiral boson's non-covariant formulation, this 6D non-covaraint action \eqref{51} for self-dual two-form also has a modified Lorentz transformation law defined by (in the gauge $B_{i5}=0$ where $H_{ij5}=\del_5 B_{ij}$.  And we should only focus on the transformations mixed with the 5-th coordinate)
\bea
\label{51l}
\delta B_{ij}&=&(\Lam\cdot x) \tilde H_{ij5}-x_5(\Lam\cdot \del)B_{ij}\nn\\
&=&(\Lam\cdot x) \del_5 B_{ij}-x_5(\Lam\cdot \del)B_{ij}-(\Lam\cdot x)(H_{ij5}+{1\over 6}\eps^{ijnpq}H_{npq}) \ . 
\eea  
One finds again that the standard Lorentz transformation law is modified by the self-duality condition. It is instructive to examine this symmetry explicitly. We have
\bea
\delta S_{5+1}&\sim& \int d^6x \delta B_{ij} \eps^{ijklm}\del_k (\tilde H_{lm5}-\del_5 B_{lm})\nn\\
&=& \int d^6x \Big((\Lam\cdot x) \tilde H_{ij5}-x_5(\Lam\cdot \del)B_{ij}\Big) \eps^{ijklm}\del_k (\tilde H_{lm5}-\del_5 B_{lm}) \ .
\eea 
So there are four terms and we write them explicitly as follows:
\bea
(1)&&= \int d^6x  (\Lam\cdot x) \tilde H_{ij5} \eps^{ijklm}\del_k (\tilde H_{lm5})\nn\\
&&=\int d^6x   [-{1\over 2}\eps^{ijklm} \Lam_k \tilde H_{ij5} (\tilde H_{lm5})]+tot.\nn\\
&&= \int d^6x  [-\Lam_k \tilde H_{ij5}H^{ijk}]+tot\nn\\
&&= \int d^6x [ -(\Lam\cdot \del)B_{ij} \tilde H^{ij5}]+tot.
\eea
\bea
(2)&&=\int d^6x  [-(\Lam\cdot x) \tilde H_{ij5} \eps^{ijklm}\del_k \del_5 B_{lm}]\nn\\
&&\sim \int d^6x (\Lam\cdot x)\tilde H_{ij5} \del_5 \tilde H^{ij5}\nn\\
&&=tot.
\eea
\bea
(3)&&=\int d^6x [ -x_5(\Lam\cdot \del)B_{ij} \eps^{ijklm}\del_k \tilde H_{lm5}]\nn\\
&&\sim\int d^6x  [x_5(\Lam\cdot \del)B_{ij} \del_k H^{ijk}]\nn\\
&&\sim\int d^6x [x_5(\Lam\cdot \del)\del_kB_{ij} H^{ijk}]+tot.\nn\\
&&\sim \int d^6x  [x_5(\Lam\cdot \del)H_{ijk} H^{ijk}]+tot.\nn\\
&&= tot.
\eea
\bea
(4)&&=\int d^6x [ x_5(\Lam\cdot \del)B_{ij} \eps^{ijklm}\del_k (\del_5 B_{lm})]\nn\\
&&=\int d^6x [-{1\over 2}(\Lam\cdot \del)B_{ij} \eps^{ijklm}\del_k B_{lm}]+tot.\nn\\
&&=\int d^6x [ (\Lam\cdot \del)B_{ij} \tilde H^{ij5}]+tot.
\eea Thus we find that, up to total derivatives, the action is invariant under the modified Lorentz transformation \eqref{51l}.
\\

Next we turn to the crucial question that how to obtain the self-duality condition from the action. We start from the equation of motion of $B_{j5}$, which reads
\bea
\del^i\tilde H_{ij5}=0 \ .
\eea 
But this is just an identity if we recall $\eps_{ijklm} \del^i\del^k=0$. This result implies that component $B_{i5}$ only enters the action through a surface term. It means that we have an addtional gauge symmetry in the theory
\bea
\delta B_{i5}=\Phi_{i} \label{agau} \ ,
\eea 
for an arbitrary one-form gauge parameter $\Phi_{i}$. On the other hand, the equation of motion of $B_{ij}$ is given by
\bea
\del^k{\cal H}_{ijk}=0
\eea 
whose solution is written as
\bea
{\cal H}_{ijk}=\eps_{ijklm}\del^l \Psi^m
\eea
for an arbitrary one-form gauge parameter $\Psi_{m}$. Now we see that under the gauge transformation (\ref{agau}) one can absorb $\Psi_i$ by using the gauge parameter $\Phi_i$. We then  obtian the self-duality condition of 2-form potentials
\bea
{\cal H}_{ijk}=0 \ .
\eea 
This result implies ${\cal H}_{ij5}=0$. This is the only self-duality condition that we need in the 5+1 splitting non-covaraint formulation.
\\

Notice that we have picked 5-th as our special direction and we used the gauge $B_{i5}=0$ above. If we integrate over the 5-th direction on both side of the self-duality condition, we get (will not be used in later sections)
\bea
\int d x^5 (-{1\over 6} \eps_{ijklm}H^{klm})= B_{ij}(x^5=\infty)-B_{ij} (x^5=-\infty) \ .
\eea 

\subsection{3+3 Splitting Formulation}
Another formulation of chiral 2-form is based on the 3+3 decomposition of spacetime 
\bea R^{1,5}=R^{1,2}\times R^3\ . \eea
Originally this consideration was motivatied by studying Bagger-Lambert-Gustavsson (BLG) multiple M2-branes model via promoting Lie 3-algebra into the infinite-dimensional symmetry of volume preserving diffeomorphisms of an internal 3-dimensional space. Physically this 3+3 splitting means that the 3-dimensional worldvolume of M2-branes can combine with extra 3-dimensional internal space to form a 6-dimensional worldvolume of a single M5-brane that carries the chiral 2-form \cite{NPM5}.
\\

The 6D Lorentz symmetry $SO(1,5)$ here is broken to $SO(1,2)\times SO(3)$ and the gauge potentials are decomposed into following three types:
\bea
B_{\m\n}=(B_{ab},~B_{a\dot b},~B_{\dot a\dot b}) \ ,
\eea 
where $a=(0,1,2)$ and $\dot a=(3,4,5)$. Notice that in this way $B_{ab}$ or $B_{\dot a\dot b}$ has three components while $B_{a\dot b}$ has nine components. The corresponding action is \cite{NPM5}
\bea
\label{33}
S_{3+3}&=&-{1\over 12} \int d^6 x [\tilde H_{a b c }(H^{a b c}-\tilde H^{a b c})+{3} \tilde H_{a  b\dot c}(H^{ a  b\dot c}-\tilde H^{ a  b\dot c})]\nn\\
&\equiv&-{1\over 12} \int d^6 x [\tilde H_{a \dot b\dot c }{\cal H}^{a\dot b \dot c}+{3} \tilde H_{ a b\dot c}{\cal H}^{a b\dot c}] \ .
\eea 
One can check that components $B_{ab}$ enter the action only through total derivatives; Its field equation is trivial (vanishes identically) and is given by
\bea
(\del^c\tilde H_{abc}+3 \del^{\dot c} \tilde H_{ab\dot c})=0 \ .
\eea 
This means there exists an additional gauge symmetry in the theory. One finds it reads
\bea
\delta B_{ab}=\Phi_{ab} \label{3ag} \ ,
\eea 
for some arbitrary 2-form gauge parameter $\Phi_{ab}$.
\\

Notice that in this 3+3 formulation, we will need to show both the following equations at the same time
\bea
{\cal H}^{a\dot b \dot c}&=&0 \ ,\\
{\cal H}^{\dot a \dot b\dot c}&=&0 \ ,
\eea to justify that the action \eqref{33} is indeed correct for describing chiral 2-forms in 6D. However, we find only a special gauge transformation $B_{ab}$ (\ref{3ag}).  Is it possible to achieve the both equations simply under a single gauge transformation?  We next see in the following that the answer is positive.
\\

The equations of motion derived from varying the action with respect to components $B_{a\dot b}$ and $B_{\dot a\dot b}$ are given by
\bea
\del^{\dot c} {\cal H}_{a\dot b\dot c}&=&0 \ ,\\
\del_{c} {H}^{\dot a\dot b c}+\del_{\dot c} {H}^{\dot a\dot b\dot c}&=&0 \ .
\eea 
We see that the first equation implies the solution
\bea
{\cal H}_{a\dot b\dot c}={1\over 2} \eps_{\dot a\dot b\dot c}\eps_{abc} \del^{\dot a}\Phi^{bc} \label{3 eom} \ ,
\eea for some arbitrary tensor $\Phi^{bc}$. After making the gauge transformation: $\delta B_{ab}=\Phi_{ab}$, one part of self-duality conditions could be obtained
\bea
{\cal H}_{a\dot b\dot c}=0 \ .
\eea  Using this result, we further rewrite another field equation as a total derivative 
\bea
\del^{\dot c}(H_{\dot a\dot b\dot c}+{1\over 2} \eps_{\dot a\dot b\dot c}\eps_{abc} \del^a B^{bc})=0 \ .
\eea We then solve it by considering
\bea
H_{\dot a\dot b\dot c}+{1\over 2} \eps_{\dot a\dot b\dot c}\eps_{abc} \del^a B^{bc}=\eps_{\dot a\dot b\dot c}\Psi(x_a) \ ,
\eea for some function $\Psi(x_a)$ independent of coordinates $x_{\dot a}$.  The next trick is that we could still redefine
\bea
B_{ab}\rightarrow B_{ab}+{1\over 3}\eps_{abc} \Phi^c(x_a) \ ,
\eea where $\Phi^c (x_a)$ is selected as $\del_c \Phi^c(x_a)=\Psi (x_a)$. (It is important to notice that since $\Phi^c (x_a)$ is independent of coordinates $x_{\dot a}$, this further redefinition/transformation of $B_{ab}$ does not spoil another self-duality condition.) We then finaly arrive the remaining self-duality condition
\bea
{\cal H}_{\dot a \dot b\dot c}&=&0 \ .
\eea 

Let us conclude this subsection by mentioning  the corresponding modified Lorentz transformation laws for this 3+3 formulation. First of all, it is obvious that the action is manifestly invariant under the $SO(1,2)\times SO(3)$ subgroup of the full $SO(1,5)$ Lorentz group. The Lorentz symmetries mixing $x_a$ and $x_{\dot a}$ are no longer manifest. The claim is that the action is still invariant under the following modified Lorentz transformation laws that are parametrized by $3\times3$ constant matrix $\lam_{a\dot a}$
\bea
\delta B^{a\dot a}&=&\lam^{a}_{\dot b} B^{\dot b \dot a}+\lam^{b}_{\dot c}(x_b \del^{\dot c}-x^{\dot c}\del_b)B^{a\dot a}+\lam^{\dot d}_{c} x_{\dot d} {\cal H}^{ca\dot a} \ ,\\
\delta B^{\dot a\dot b}&=&\lam^{\dot b}_{ a} B^{a \dot a}-\lam^{\dot a}_{a} B^{a \dot b}+\lam^{b}_{\dot c}(x_b \del^{\dot c}-x^{\dot c}\del_b)B^{\dot a\dot b} \ .
\eea Here we have considered the gauge $B_{ab}=0$ again. One sees that these modified transformation laws become the standard Lorentz transformation on  shell.

\subsection{4+2 Splitting Formulation}
We have just studied how to construct 6D chiral 2-form theories based on $5+1$ splitting and $3 + 3$ splitting. It is now nature to ask for a formulation based on 4+2 splitting. This is given by \cite{Chen:2010jgb}.  Here the six dimensional Lorentz symmetry $SO(1,5)$ is broken to $SO(1,1)\times SO(4)$. In this formulation, one considers the 4+2 decomposition of spacetime
\bea
R^{1,5}=R^{1,3}\times R^2 \ . 
\eea The gauge 2-form potentials are decomposed into
\bea
B_{\m\n}=(B_{ab},~B_{a\dot b},~B_{\dot a\dot b}) \ ,
\eea where $a=(0,1)$ and $\dot a=(2, 3,4,5)$. The action is given by  \cite{Chen:2010jgb}
\bea
S_{2+4}=-{1\over 8}\int d^6 x \Big(2\tilde H_{ab\dot c}
{\cal H}^{ab\dot c}+\tilde H_{a\dot b\dot c}{\cal H}^{a\dot b\dot c}\Big).
\eea
The field equation of $B_{ab}$ is again trivial (vanishes identically), and is give by
\bea
\partial^{\dot a}\tilde H_{ab\dot a}= 0,
\eea
that implies that we have the following gauge transformation
\bea
\delta B_{ab}=\Phi_{ab}.
\eea
The field equations of $B_{a\dot a}$ and $B_{\dot a\dot b}$ are
\bea
\partial^{\dot b}{\cal H}_{a\dot a\dot b}&=&0\label{4eom1} \ , \\
\partial^{\dot c}({\cal H}+H)_{\dot a\dot b\dot c}+\partial^a H_{a\dot a\dot b}&=&0.\label{4so2}
\eea
The solution to the first equation (\ref{4eom1}) could be written as
\bea
{\cal H}_{a\dot a\dot b}=\epsilon_{ab}\epsilon_{\dot a\dot b\dot c\dot d}\partial^{\dot c}\Psi^{b	\dot d} \label{4so}
\eea
for some arbitrary functions $\Psi^{b \dot d}$. By taking the Hodge-dual of both sides, we now get
\bea
{\cal H}_{a\dot a\dot b}=\partial_{\dot a}\Psi_{a\dot b}-\partial_{\dot b}\Psi_{a\dot a}.
\eea
Identifying these two results, we have
\bea
\partial_{\dot a}\Psi_{a\dot b}-\partial_{\dot b}\Psi_{a\dot a}=
\epsilon_{ab}\epsilon_{\dot a\dot b\dot c\dot d}\partial^{\dot c}\Psi^{b	\dot d},
\eea
Now we act $\partial^{\dot b}$ on both sides of the equivalence relation to get
\bea
\label{phi4}
\partial^{\dot{b}}\partial_{\dot{b}}\Psi_{a\dot a}-\partial_{\dot a}\partial^{\dot b}\Psi_{a\dot b}=0.
\eea Because of the solution (\ref{4so}) is unchanged
under the transformation
\bea
\Psi_{a\dot a}\rightarrow \Psi_{a\dot a}+\partial_{\dot a}\Lam_a \ ,
\eea
we choose the Lorenz gauge to fix that redundant symmetry, i.e.
$\partial^{\dot a}\Psi_{a\dot a}=0$. Now \eqref{phi4} reduces to
\bea
\dot \partial^2\Psi_{a\dot a}=0,~(
\dot \partial^2\equiv \partial^{\dot a}\partial_{\dot a}).
\eea
Next we consider a proper boundary condition that $\Psi^{a\dot a}$ vanishes at infinities of the 4D Euclidean space with coordinates $x_{\dot a}$
such that it has the unique solution
$\Psi_{a\dot a}=0$.  We obtain one of the self-duality conditions
\bea
{\cal H}_{a\dot a\dot b}=0.
\eea 

If we plug this result into the equation of motion of $B_{\dot a\dot b}$ (\ref{4so2}),  we get the solution 
\bea
{\cal H}_{\dot a\dot b\dot c}=\frac{1}{2}\epsilon_{ab}\epsilon_{\dot a\dot b\dot c\dot d}\partial^{\dot d}\phi^{ab} \ ,
\eea
for some arbitrary functions $\phi^{ab}$. Now we use the gauge transformation $\delta B_{ab}=\Phi_{ab}$ so that this equation becomes 
\bea
{\cal H}_{\dot a\dot b\dot c}=0.
\eea
So we obtain self-duality conditions for all components of the field strengths in this 4+2 formulation.
\\

Let us summarize this subsection by providing the modified Lorentz transformation laws of this 4+2 formulation. The action is manifestly invariant under the $SO(1,1)\times SO(4)$ subgroup of the full $SO(1,5)$ Lorentz group, while the Lorentz symmetries mixing $x_a$ and $x_{\dot a}$ are not manifest. The modified Lorentz transformation laws that are parametrized by $2\times 4$ constant matrix $\lam_{a\dot a}$ are  
\bea
\delta B^{a\dot a}&=&\lam^{a}_{\dot b} B^{\dot b \dot a}+\lam^{b}_{\dot c}(x_b \del^{\dot c}-x^{\dot c}\del_b)B^{a\dot a}+\lam^{\dot d}_{c} x_{\dot d} {\cal H}^{ca\dot a}\ ,\\
\delta B^{\dot a\dot b}&=&\lam^{\dot a}_{a} B^{\dot b a}-\lam^{\dot b}_{ b} B^{\dot a b}+\lam^{b}_{\dot c}(x_b \del^{\dot c}-x^{\dot c}\del_b)B^{\dot a\dot b} \ .
\eea We here consider the gauge $B_{ab}=0$ for simplicity. The modified transformation laws become the standard Lorentz transformation on  shell.

\newpage
\section{Covariant Formulations: PST Model}
We have constructed several non-manifest Lorentz invariant actions for self-dual fields. However, it is always desirable to find covariant formulations, especially when one wants to consider more complicated cases. For example, the case when fields couple to gravity. It turns out that we need to introduce auxiliary fields to construct covariant self-dual field theories. In fact, people have tried to consider different numbers of auxiliary field that vary from infinite auxilirary fields to only one auxiliary field. We here consider the most modern approach, called PST formulation \cite{PST}.
\\

The general lesson we will learn is that there will be an extra gauge symmetry that allows us to gauge fix the auxiliary field to an proper configuration and the  PST covariant formulations will simply reduce to previous non-covariant self-dual (PS) actions. We will again start from the simplest case: A covariant action  of the 2D chiral boson.  We next consider covariant formulation of the 6D self-dual two-form theory for a single M5-brane.

\subsection{Covariant action for Chiral Boson: D=2}
The PST covariant action for a two–dimensional chiral boson was constructed by using a single auxiliary scalar field denoted by $a (x)$. This auxiliary field should be spacetime dependent for the sake of having the manifest Lorentz symmetry. The 2D action is given by \cite{PST}
\bea
\label{2dpst}
S=\int d^2 x \Big[ -{1\over 2} F^a F_a+{1\over 2({\del a})^2} (\del^b a {\cal F}_b)^2 \Big] \label{1p} \ ,
\eea where $a,b =0,1$. Here we will not need to distinguish time or space coordinate explicitly. Define $F_a=\del_a\phi
$ as the field strength of the boson $\phi(x)$ as before and ${\cal F}_a$ is defined by 
\bea 
{\cal F}_a=F_a-\eps_{ab} F^b  \ ,
\eea as the expression for the self-duality condition. One can check that the action \eqref{2dpst} has the following two sets of gauge transformation
\bea
&(1)& \delta \phi= f(a),~\delta a=0 \ , \\ \label{2g}
&(2)& \delta \phi= {\psi\over ({\del a})^2} \del^b a {\cal F}_b,~ \delta a=\psi(x) \ , \label{2gg}
\eea 
defined by gauge parameters $f(a)$ and $\psi(x)$. The second gauge symmetry plays an important role that allows us to guage-fix the auxiliary field to, for example, $a=x$. This gauge-fixing choice reduces the 2D PST covariant action \eqref{2dpst}  to the previous non-covaraint 2D FJ action \eqref{FJJ}.
\\

Next let us see how to obtain the self-duality condition from this covariant action. The equation of motion of $\phi(x)$ is given by
\bea
\eps^{ab}\del_b[{1\over 2({\del a})^2} \del_a a\del_c a {\cal F}^c]=0 \ ,
\eea 
which implies the following solution
\bea
{1\over ({\del a})^2} \del_a a\del_c a{\cal F}^c=\del_a\Omega \ ,
\eea 
for some arbitrary scalar $\Omega$. 
We can project this solution with two orthogonal vectors: $\del_a a$ and $\eps^{ab} \del_b a$, to get the following two equations
\bea
{1\over 2} \del_c a {\cal F}^c&=&\del^a a\del_a\Omega \label{1pst} \ , \\
\eps^{ab} \del_a a\del_b\Omega&=&0 \label{2pst} \ .
\eea 
We see the solution to the equation (\ref{2pst}) can be given by
\bea
\Omega=\tilde f(a) \ ,
\eea 
for some arbitrary function of $a(x)$. Plugging it into the first equation (\ref {1pst}), we obtain
\bea
{1\over 2} \del_c a {\cal F}^c&=&\del^a a\del_a\tilde f(a) \ .
\eea 
On the other hand,  under the gauge transformation $\delta \phi= f(a),~\delta a=0$, the left hand side of the equation (\ref{1pst}) becomes
\bea
{1\over 2} \del_c a\delta {\cal F}^c={1\over 2} \del^a a\del_a f(a) \ ,
\eea 
thus we could use the freedom from parameter $f(a)$ to absorb $\tilde f(a)$. We get 
\bea
{1\over 2} \del_c a {\cal F}^c=0 \ .
\eea  Notice that this implies ${\cal F}^1=0$, if one fixes $a(x)$ to $x^1$, we have $\del_{0}\phi-\del_{1}\phi=0$, which is in fact all we need for a chiral boson. Fixing $a(x)$ to $x^0$ will give the same result. Also notice that the equation of motion of $a$ is trivial in the sense that it is proportional to the self-duality condition. 
\\

We also notice that the action \eqref{1p} shares the same form as Seigel's action \eqref{1S}. The difference is that the self-duality conditon obtained from Seigel's action is through the field equation of auxiliary field $\lambda$ while in the PST formulation, the self-duality condition is obtained from the field equation of $\phi$ with an additional gauge transformation. These two theories are the same in classical level in the sense that they have the same equation of motion.

\subsection{Covariant action for Chiral 2-form: D=6}
The Lorentz-covariant action for the 6D Abelian chiral 2-form theory (or the gauge sector for a single M5-brane) is constructed also with the help of a single auxiliary scalar field. We will see again that the corresponding PST covariant model has an extra gauge symmetry which can be used to gauge-fix the auxiliary field and the action reduces to the previous non-covariant 6D PS action \eqref{51}.\\

The covaraint action is given by \cite{PST}:
\bea
S={1\over 4} \int d^6x~\Big ( -{1\over 3!} H^{\m\n\lambda}H_{\m\n\lambda}+{1\over 2} {\cal H^{\m\n\rho}} P^\sigma_\rho {\cal H_{\m\n\sigma}}\Big) \ , \label{c51}
\eea 
where
\bea
{\cal H_{\m\n\sigma}}= (H-\tilde H)_{\m\n\sigma} \ ,
\eea
with $\tilde H$ represents the Hodge dual of H 
\bea
\tilde H^{\m\n\lambda}=-{1\over 6} \eps^{\m\n\lambda abc}H_{abc} \ .
\eea
We also denote 
\bea
P^\m_\n= {\partial^\m b \partial_\n b\over (\partial b)^2} \ ,
\eea in terms of an auxiliary field $b(x)$. Since the algebra in six-dimensions is much more complicated than the previous two-dimensional model, it would be helpful to derive equations of motion explicitly. We will later observe the crucial existence of an extra gauge symmetry.
\\

The equation of motion of $B_{\alpha\beta}$ from the first term in the action gives 
\bea
{\delta \Big(-{1\over 3!} H^{\m\n\lambda}H_{\m\n\lambda}\Big)\over \delta B_{\alpha\beta}}&=&\partial_\g H^{\alpha\beta\g}=\partial_\g \cal H^{\alpha\beta\g}\nn\\
&=&\partial_\g\Big( {-1\over 2} \eps^{\a\b\g\m\n\lam} P^\sigma_\m {\cal H_{\sigma\n\lam}}+ 3 P^{[\a}_\m {\cal H^{\b\g]\m}} \Big) \ ,
\eea
where we have considered the identity
\bea
\eps^{\a\b\g\m\n\lam} P^\sigma_\m {\cal H_{\sigma\n\lam}}&=&-{1\over 6} \eps^{\a\b\g\m\n\lam}\eps_{\sigma\n\lam\k\eta\delta}P^\sigma_\m {\cal H^{\k\eta\delta}}= {2!4!\over 6}\delta^{[\a\b\g\m]} _{[\sigma\k\eta\delta]}P^\sigma_\m {\cal H^{\k\eta\delta}}\nn\\
&=&{1\over 6} \Big(-2!3! P^{\m}_{\m}{\cal H^{\a\b\g}}+2!(4!-3!) P^{[\a}_\m {\cal H^{\b\g]\m}}\Big)\nn\\
&=& -2 {\cal H^{\a\b\g}}+ 3! P^{[\a}_\m {\cal H^{\b\g]\m}}  \ .\label{id} 
\eea
The variation of the second term gives
\bea
{\delta \Big({1\over 2} {\cal H^{\m\n\rho}} P^\sigma_\rho {\cal H_{\m\n\sigma}}\Big)\over \delta B_{\a\b}}={\Big( {\cal H^{\m\n\rho}} P^\sigma_\rho \delta {\cal H_{\m\n\sigma}}\Big)\over \delta B_{\a\b}}=-\partial_\g \Big(3 P^{[\a}_\m {\cal H^{\b\g]\m}}+ {1\over 2} \eps^{\a\b\g\m\n\lam} P^\sigma_\m {\cal H_{\sigma\n\lam}} \Big) \ .
\eea
Overall we have the variation on the action with respect to $B_{\m\n}$ given by
\bea
-\eps^{\alpha\beta\g\m\n\lambda}\partial_\m b (\partial_\g \bar H_{\n\lambda})\delta B_{\a\b} \ ,
\eea
where
\bea
\bar H^{\m\n}\equiv{\cal H^{\m\n\rho}}{ \partial_\rho b\over (\partial b)^2} \ .
\eea
Next we consider the equation of motion of $b(x)$, which only appears in the second term of the action. We calculate
\bea
{\delta \Big( {1\over 2} {\cal H^{\m\n\rho}} P^\sigma_\rho {\cal H_{\m\n\sigma}}\Big)\over \delta b}
&=&-\partial^\g\Big[{\cal H^{\m\n\rho}}{ \partial_\rho b\over (\partial b)^2}{\cal H_{\m\n\g}}\Big]+\partial^\g\Big[{\cal H^{\m\n\rho}}P^\sigma_\rho { \partial_\g b\over (\partial b)^2}{\cal H_{\m\n\sigma}}\Big]\nn\\
&=& -\partial^\g(\bar H^{\m\n} {\cal H_{\m\n\g}})+\partial^\g (\bar H^{\m\n}\bar H_{\m\n} \partial_\g b ) \ .
\eea
 We notice that the first term can be written as (using \eqref{id}):
\bea
-\partial^\g(\bar H^{\a\b} {\cal H_{\a\b\g}})&=&-\partial_\g(\bar H_{\a\b} {{-1\over 2} \eps^{\a\b\g\m\n\lam} P^\sigma_\m {\cal H_{\sigma\n\lam}}+ 3 \bar H_{\a\b}P^{[\a}_\m {\cal H^{\b\g]\m}}})\nn\\
&=&-\partial_\g(\bar H_{\a\b} {{-1\over 2} \eps^{\a\b\g\m\n\lam} \bar H_{\n\lambda}\partial_\m b+ \bar H_{\a\b}P^{\g}_\m {\cal H^{\a\b\m}}}) \ ,
\eea where we have used the fact that
\bea
\bar H_{\m\n} \partial^\n b={\cal H_{\m\n\lambda}} {\partial^\lambda b \partial^\n b\over (\partial b)^2}=0 \ .
\eea
Overall we have the variation of the action with respect to $b(x)$ given by
\bea
{1\over 2} \partial_\g \Big( \bar H_{\a\b}\eps^{\a\b\g\m\n\lam} \bar H_{\n\lam}\partial_\m b \Big)\delta b=
\bar H_{\a\b}\eps^{\a\b\g\m\n\lam} \Big(\partial_\g\bar H_{\n\lam}\Big)\partial_\m b ~\delta b \ .
\eea
In sum, the variation of the ''Abelian'' PST action gives
\bea
\label{vv}
\delta S=\int d^6 x\Big( -\eps^{\alpha\beta\rho\m\n\lambda}\partial_\m b (\partial_\rho \bar H_{\n\lambda}) (\delta B_{\alpha\beta})+\eps^{\alpha\beta\rho\m\n\lambda}~\partial_\m b (\partial_\rho \bar H_{\n\lambda} )\bar H_{\alpha\beta} ~(\delta b)\Big) \ . \label{sum}
\eea
so it is obvious we have an extra symmetry under the gauge parameter $\phi$
\bea
\delta B_{\m\n}&=& \bar H_{\m\n} \phi \ , \\
\delta b &=&\phi \ .
\eea This set of new gauge symmetry allows us to gauge-fix the field $b(x)$ to $x^5$ and the covariant action redueces to the non-covariant action \eqref{51}. Notice a subtle point: One should avoid setting $b=0$ since it brings the singularity into the theory as field $b$ appears in the denominator.
\\

The next question is how to obtain the self-duality condition from this covariant action. We achieve this goal by observing that the general solution of the equation of motion of $B_{\m\n}$ (one can simply read it from the first term in \eqref{sum}) can be given by letting
\bea
{\cal H}_{\m\n\lam}\del^\lam b= (\del b)^2 \del_{[\m} \Phi_{\n]}+\del_{[\m} b \del_{\n]} \Phi_\rho \del^\rho b+\del^\rho b \del_\rho \Phi_{[\m} \del_{\n]} b \ , \label{csd}
\eea 
for some arbitrary vector $\Phi_\m$.  Next, just like in the previous D=2 case, we should be able to find a way to ''absord'' the right hand side of the above equation to obtain the self-duality condition.  Indeed, there is an additional gauge symmetry of the action, and the transformation is give in terms of the gauge parameter $\Phi_\m$
\bea
\delta B_{\m\n}&=&(\del_\m b) \Phi_\n-(\del_\n b) \Phi_\m \label{ex}\ , \\
\delta b&=&0 \ .
\eea 
This symmetry will generate the same form as the right hand side in \eqref{csd} when considering $\delta ({\cal H}_{\m\n\lam}\del^\lam b)$. So now we obtain the self-duality condition for the 2-form potential. 
\\

One can also check that the equation of motion for the auxiliary field $b$ is trivial because it is proportional to the self-duality condition. Also it is interesting notice that if we gauge-fix the auxiliary field by
$a=x^5$,
the extra gauge symmetry of $B$ \eqref{ex} reduces to $\delta B_{i5}=\Phi_{i5} $, which is just the extra gauge symmetry \eqref{agau} used to obtain the self-duality condition in the previous non-covaraint PS action.

\newpage
\section{Non-Abelain Chiral 2-Form: Multiple M5-Branes}
Multiple M5-branes system has been the most challenging and mysterious brane system in string/M theory. A single M5-brane is described by Abelian self-dual 2-form gauge field theory that we have discussed in the previous sections. When branes start to close to each other, interaction among M5-branes will appear and we expect multiple M5-brane  should be described by a certaion ''non-Abelian'' self-dual 2-form gauge field theory.  As we have seen in previous sections, to construct an action for self-dual fields is already a highly non-trivial task, here we will face a even bigger challenging: To formulate a consistent non-Abelian 2-form theory. 
\\

Notice that formulating a non-Abelian 2-form gauge theory should be totally independent from formulating a self-dual field theory. Only having non-Abelian 2-form gauge theory is not sufficient to describe M5-branes. We will need to combine these two characteristics in the end. Let us firstly focus on non-Abelian 2-form potentials without the self-duality.

\subsection {Non-Locality}
Several no-go theorems \cite{no-go} have claimed that such an exotic system (multiple M5-branes) is impossible to build. They considered general deformations of N commuting copies of Abelian self-dual 2-form gauge theory in 6 dimensions, and they found that a consistent non-Abelian deformation is always trivial in the sense that it will be equivalent to simply a change of variables that does not really deform the gauge algebra. However, every no-go theorem has their weakness, although they did not assume the deformed gauge transformations to correspond to a particular type of symmetry algebra, they did assume that the deformation was ${local}$. But accepting the non-locality of M5-branes theory is not enough to see how to construct such a theory and we certainly need some other ideas.  
\\

We will be constructing a non-Abelian self-dual 2-form gauge theory in 6 dimensions with a spatial direction compactified on a circle of radius R. (The reason for the compactification will be given later.) It has the following two important properties to justify that it can be a correct theory for M5-branes. (1) It reduces to the Yang-Mills theory (multiple D4-branes) in 5 dimensions for small radius R. (2) It is equivalent to the Lorentz-invariant theory of Abelian chiral 2-forms when we turn off the coupling. Previous no-go theorems prohibiting non-Abelian deformations are circumvented by introducing nonlocality along the compactified dimension.\cite{HHM}

\subsection{Main Ideas}
The main ideas of this formulation are largely two-fold. First, although there is no rule telling us that we must introduce covariant derivatives in this kind of non-abelian higher form gauge theory. However, it is unclear how to have a gauge covariant structure without introducing covaraint derivatives. Also notice that if we have a non-Abelian 2-form theory, we expect that after compactification, the gauge symmetry structure should fully reduce to the non-Abelian 1-form theory, which is just the standard Yang-Mills theory.  In Yang-Mills theory, we use the covariant derivatives and the gauge transformation is 
\bea
\delta A_\m=[D_\m, \lambda] \ ,
\eea 
in terms of the 0-form gauge parameter $\lambda$. In short, it seems still better that we can have covaraint derivatives. But there should be a 1-form potential together with ordinary derivatives to define covaraint derivatives. Here comes the first question: What is this 1-form in the theory of M5-branes?
\\

Naively, one might try to introduce a one-form potential $A_{\m}$. However, in the theory of M5-brane(s), there is no such field. Introducing extra field potentially implies having more physical degrees of freedom. One may try to consider this $A_{\m}$ as an auxiliary field, for example it could be gauged away by an extra gauge symmetry. But if it is the case, the theory is the same as a theory without $A_\m$. On the other hand, Chern-Simon actions are only defined in odd dimensions, it may be possible to find a certian kind of BF-like topological theory using the one-form potential in six dimensions to construct the covariant derivatives. But it is also not clear how to do it. 
\\

In this formulation, we consider compactified M5-branes on a circle. The worldvolume of M5-branes is on 
\bea
R^{1,4}\times S^1 \ ,
\eea where $S^1$ is a circle with the radius R  with a periodic coordinate 
\bea x^5\sim x^5+2\pi R  \ .
\eea  Here we consider a space-like circle for the sake of reducing to D4-branes theory in $R^{1,4}$.  This consideration allows us to define covaraint derivatives in terms of the {$zero$ $modes$ of $B_{i5}$}
\bea
D_i=\del_i+gB^{(0)}_{i5} \ .
\eea Note that the zero modes of $B^{(0)}_{i5}$ is just the Yang-Mills one-form vector potential, thus covariant derivatives we use is the same in the standard Yang-Mills theory. 
\\

For an arbitrary field $\Phi$, we have the decomposition
\bea
\Phi(x_\m)&=&\Phi^{(0)}(x_i)+ \sum_{(n)} \Phi_n(x_i) e^{i{n\over R}x_5} \\
&\equiv&\Phi^{(0)}+\Phi^{(KK)}
\eea
where we use the notation that the superscript ''(0)'' represents zero modes while ''(KK)'' represents Kaluza-Klein modes. Obviously, we have
\bea
\del_5 \Phi^{(0)}&=&0\ , \\
\del_5 \Phi&=&\del_5\Phi^{(KK)} \ .
\eea

The second key idea in this model is that we will introduce the non-locality in the theory through defining a non-local operator
\bea \del^{-1}_5 \ .
\eea
This operator consistently acts on KK modes: When $\del_5$ acts on KK modes, it gives $\sim{n\over R}$ with nonzero $n$ so that this operator is invertible. And this operator satisfies
\bea
\del^{-1}_5 \del_5 \Phi= \Phi^{(KK)} \ .
\eea 
The non-locality that we introduce could be considered as the way to circumvent previous no-go theorems \cite{no-go} claiming that it is impossible to construct a theory of M5-branes, as those theorems are all based on the assumption of the locality in the theory. We will see this non-local operator indeed plays an important role in order to have a consistent formulation.
\subsection{Non-Abelian Gauge Transformations}

The non-Abelian generalization of gauge transformation laws of the anti-symmetry tensor field (2-form) $B_{ij}$ are found to be \cite{HHM}
\bea
\delta B_{i5}&=& [D_i,\Lambda_5]-\del_5\Lam_i+g[B^{(KK)}_{i5},\Lam^{(0)}_5]\       
, \\
\delta B_{ij}&=& [D_i,\Lambda_j]-[D_j,\Lam_i]+g[B_{ij},\Lam^{(0)}_5]-g[F_{ij},\del^{-1}_5 \Lam^{(KK)}_5] \ , 
\eea 
where $i,j=0,1,2,3,4.$ and parameter $g$ is the coupling constant with the mass dimension. (Note that in the M-theory, there is no adjustable parameter, we will see later that the coupling constant $g$ turns out to be the radius R of the M-circle.)  The covaraint derivative is given in terms of zero modes $D_i=\del_i+gB^{(0)}_{i5}$ and the 2-form field strength $F$ is defined via the standard way
\bea
F_{ij}=g^{-1}[D_i,D_j]=\del_i B^{(0)}_{j5}-\del_j B^{(0)}_{i5}+g[B^{(0)}_{i5},B^{(0)}_{j5}] \ .
\eea 
Since $F_{ij}$ is pure zero modes, we have $\del_5 F_{ij}=0$.  More explicitly, we can decompose the gauge transformation laws into
\bea
\delta B^{(0)}_{i5}&=&[D_i, \Lam^{(0)}_{5}] \ ,\\
\delta B^{(KK)}_{i5}&=&[D_i,\Lam^{(KK)}_5]-\del_5\Lam^{(KK)}_i+g[B^{(KK)}_{i5},\Lam^{(0)}_5]\ , \\
\delta B^{(0)}_{ij}&=&[D_i,\Lam^{(0)}_j]-[D_j,\Lam^{(0)}_i]+g[B^{(0)}_{ij},\Lam^{(0)}_5]\ , \\
\delta B^{(KK)}_{ij}&=&[D_i,\Lam^{(KK)}_j]-[D_j,\Lam^{(KK)}_i]+g[B^{(KK)}_{ij},\Lam^{(0)}_5]-g[F_{ij},\del^{-1}_5 \Lam^{(KK)}_5] \ ,  \label{4}\nn\\
\eea 
with all quantities $B_{i5},B_{ij}, \Lam_5,\Lam_{i}$ take values in a Lie algebra G. 
One can check the algebra of gauge transformation is closed and given by
\bea
[\delta,\delta']=\delta'' \ ,
\eea with
\bea
\Lam''^{(0)}_5&=&g[\Lam^{(0)}_5,\Lam'^{(0)}_5]\ , \\
\Lam''^{(KK)}_5&=&g[\Lam^{(0)}_5,\Lam'^{(KK)}_5]-g[\Lam'^{(0)}_5,\Lam^{(KK)}_5] \ ,\\
\Lam''_i &=&g[\Lam^{(0)}_5,\Lam'_i]-g[\Lam'^{(0)}_5,\Lam_i] \ .
\eea
\\

Let us give some comments on the non-Abelian gauge transformation laws.
\\

First we notice that there is a non-local term in $\delta B^{(KK)}_{ij}$ \eqref{4}. Why we need that? Recall that in six dimensional spacetime, there should be only 5 independent gauge parameters for the 2-form potential gauge theory rather than 6, because of the crucial redundant gauge symmetry.  In non-Abelian case, the "gauge symmetry of the gauge symmetry'' is defined by 
\bea
\delta \Lam^{(KK)}_i&=&[D_i, \lam^{(KK)}] \ ,\\
\delta \Lam^{(KK)}_5&=&\del_5 \lam^{(KK)} \ .
\eea 
One can check that \eqref{4} is indeed invaraint under these redundant transformations. The existence of this gauge symemtry is important for this non-Abelian gauge transformation laws to be justified as the correct deformation of the Abelian gauge transformation laws of 2-form potentials.
\\

Now we can use the redundant symmetry to gauge-fix 
\bea
\Lam^{(KK)}_5=0 \ ,
\eea 
using the freedom from $\lam^{(KK)}$. The resulting guage transformation laws are equivalent to
\bea
\delta B^{(KK)}_{i5}&=&-\del_5 \tilde\Lam^{(KK)}_i+g[B^{(KK)}_{i5},\Lam^{(0)}_5] \ , \\
\delta B^{(KK}_{ij}&=& [D_i,\tilde\Lam^{(KK)}_j]-[D_j\tilde\Lam^{(KK)}_i]+g[B^{(KK)}_{ij},\Lam^{(0)}_5] \ ,
\eea 
where
\bea
\tilde\Lam^{(KK)}_i= \Lam^{(KK)}_i-[D_i,\del^{-1}_5 \Lam^{(KK)}_5] \ .
\eea
Notice that the zero modes $\Lam^{(0)}_5$ can not be gauged away since $\del_5 \lam^{(0)}=0$. (If the zero mode $\lam^{(0)}$ can be gauged away, then there is no gauge parameter left for the Yang-Mills theory of D4-branes after the dimensional reduction.) In non-Abelianizing the gauge transformations of a 2-form potential,
the zero mode $\Lam_5^{\z}$ plays a special role.
We associate the special role played by $\Lam_5^{\z}$ 
to its topological nature:
While $\Lam_5^{\nz}$ can be gauged away,
the zero mode $\Lam_5^{\z}$ corresponds to the Wilson line degrees of freedom  for the gauge transformation parameter $\Lambda$
along the circle in the $x^5$ direction.
\\

One might ask how about the gauge parameter $\Lam^{(0)}_i$ and its redundent symemtry? The fact is that we will not use $B^{(0)}_{ij}$ and hence neither $\Lam^{(0)}_i$ in this formulation, we will soon see the reason of it.  Also notice that the only non-local term in the gauge transformation laws is gauged way through the change of variables, but it does not mean that this theory is local since the non-locality is hidden in the gauge parameters we redefined.  On the other hand, we have ''additional non-locality'' introduced in the theory in the sense that the gauge transformation laws are defined separately for the zero modes and KK modes: For instance, one may think the term $\delta B^{(0)}_{i5}$ could be deduced form $\delta B^{(KK)}_{i5}$. But we see there are in fact different by a factor 2 through the commutator term. Also notice that we did not introduce terms like 
\bea
[B^{(KK)}_{i5}, \Lam^{(KK)}_5] \ .
\eea
In particular, thoughout this formulation all the commutators should involve at most one KK mode. Future works about the geometric interpretation of this formulamation might help us to better understand these interesting consequences. Finally, one could easily observe that if G is abelian, this non-Abelian guage transformation laws are reduced to the conventional Abelian gauge transformation of 2-form gauge potential, which is used to describe a single M5-brane.

\subsection{Non-Abelian Field Strengths}

We define 3-form field strengths as
\bea
H^{(0)}_{ij5} &=& F_{ij}=g^{-1}[D_i, D_j] \ ,\\
H^{(KK)}_{ij5} &=& [D_i, B^{(KK)}_{j5}]-[D_j, B^{(KK)}_{i5}]+\del_5 B^{(KK)}_{ij}\ , \\
H^{(0)}_{ijk} &=& [D_i, B^{(0)}_{jk}]+[D_j, B^{(0)}_{ki}]+[D_k, B^{(0)}_{ij}]\ , \\
H^{(KK)}_{ijk} &=& [D_i, B^{(KK)}_{jk}]+[D_j, B^{(KK)}_{ki}]+[D_k, B^{(KK)}_{ij}]\nn\\
 &&+g[F_{ij},\del^{-1}_5B^{(KK)}_{k5}]+g[F_{jk},\del^{-1}_5B^{(KK)}_{i5}]+g[F_{ki},\del^{-1}_5B^{(KK)}_{j5}] \ .
\eea 
They satisfy the generalized Jacobi identities
\bea
\sum_{(3)} [D_i,H^{(0)}_{jk5}]&=&0 \ ,\\
\sum_{(3)}[D_i,H^{(KK)}_{jk5}]&=& \del_5 H^{(KK)}_{ijk} \ , \\
\sum_{(4)}[D_i,H^{(0)}_{jkl}]&=&\sum_{(6)} [F_{ij}, B^{(0)}_{kl}]\ , \\
\sum_{(4)}[D_i,H^{(KK)}_{jkl}]&=&\sum_{(6)} [F_{ij}, \del^{-1}_5H^{(KK)}_{kl5}] \ ,
\eea where $\sum_{(n)}$ represents a sum over n terms that totally anti-symmetrized all the indices.
\\

These 3-form field strengths transform as
\bea
\delta H^{(0)}_{ij5}&=&g[H^{(0)}_{ij5},\Lam^{(0)}_5]\\
\delta H^{(KK)}_{ij5}&=&g[H^{(KK)}_{ij5},\Lam^{(0)}_5]\\
\delta H^{(0)}_{ijk}&=&g[H^{(0)}_{ij5},\Lam^{(0)}_5]+g[F_{ij},\Lam^{(0)}_k]+g[F_{jk},\Lam^{(0)}_i]+g[F_{ki},\Lam^{(0)}_j] \label{w}\\
\delta H^{(KK)}_{ijk}&=&g[H^{(KK)}_{ijk},\Lam^{(0)}_5]
\eea 
a potential problem here is that the transformation of $H^{(0)}_{ijk}$ is not covaraint. 
It causes a trouble to construct an gauge invaraint action. We will address this problem later.
\\

Let us consider a useful gauge
\bea
B^{(KK)}_{i5}=0 \ ,
\eea by using the freedom from the gauge paramter $\Lam^{(KK)}_{i5}$. 
We use $\hat B_{ij}$ as our variable to denote the theory under this gauge. We have
\bea
H^{(KK)}_{ij5} &=&\del_5 \hat B^{(KK)}_{ij} \ , \\
H^{(KK)}_{ijk} &=& [D_i, \hat B^{(KK)}_{jk}]+[D_j, \hat B^{(KK)}_{ki}]+[D_k, \hat B^{(KK)}_{ij}] \ .
\eea 
We can also wite 
\bea
\hat B_{ij}=\del_5^{-1} H^{(KK)}_{ij5} \ ,
\eea 
which transforms covaraintly: $\delta \hat B_{ij}=[\hat B_{ij},\Lam^{(0)}_5]$. This variable will be useful in later computation.

\subsection{Coupling to Antisymmetry Tensors}
Apart from multiple M5-branes,
let us also consider applications to this formulaiton. Readers who want to focus on the thoery of M5-branes might simply skip this subsection.
\\

A straightforward generalization of the transformation laws for $H$ leads to the definition of gauge transformations of 
a totally antisymmetrized tensor field $\phi_{\mu_1\cdots \mu_n}$
($n \leq 6$). We define their gauge transformation laws as
\bea
\d \phi_{i_1\cdots i_{n-1} 5}^{0} 
&=& g[\phi_{i_1\cdots i_{n-1} 5}^{(0)}, \Lam_5^{(0)}], \\
\d \phi_{i_1 \cdots i_{n-1} 5}^{\nz} 
&=& g[\phi_{i_1 \cdots i_{n-1} 5}^{(KK)}, \Lam_5^{(0)}], \\
\d \phi_{i_1\cdots i_n}^{(0)} 
&=& g[\phi_{i_1\cdots i_n}^{(0)}, \Lam_5^{(0)}]+ g\sum_{(n)} [\phi_{i_1\cdots i_{n-1} 5}^{(0)}, \Lam_{i_n}^{(0)}] \ ,\\
\d \phi_{i_1\cdots i_n}^{(KK)} 
&=& g[\phi_{i_1\cdots i_n}^{(KK)}, \Lam_5^{(KK)}],
\eea
where $\sum_{(n)}$ represents a sum of $n$ terms that totally antisymmetrizes all indices.
\\

We see that the gauge transformation law for the component $\phi_{i_1\cdots i_n}^{(o)}$
is different from all other components. It is defined to mimic the gauge transformation of $H_{ijk}^{(0)}$.
We should check whether this complication will prevent us from constructing a gauge field theory. First, products of these fields $\phi_{i_1\cdots i_n}^{(0)}$ will also transform 
in the form of not being covariant (where we consider that all indices are antisymmetrized on the products). Secondly, 
although $D_i$ acts on $\phi_{i_1\cdots i_n}^{(0)}$ does not transform covariantly, we can define a covariant exterior derivative for $\phi_{i_1\cdots i_n}^{(0)}$ as
\be
({\cal D}\phi)_{i_1\cdots i_{n+1}}^{(0)} \equiv
\sum_{(n+1)} [D_{i_{1}}, \phi_{i_2\cdots i_{n+1}}^{(0)}]
- (-1)^n \sum_{((n+1)n/2)} [B_{i_1 i_2}^{(0)}, \phi_{i_3\cdots i_{n+1} 5}^{(0)}].
\ee
(This expression is nontrivial only if $n \leq 5$.)
This covariant exterior derivative is indeed covariant, that is,
\be
\d ({\cal D}\phi)_{i_1\cdots i_{n+1}}^{(0)} =
g[({\cal D}\phi)_{i_1\cdots i_{n+1}}^{\z}, \Lam_5^{(0)}]
+ g\sum_{(n+1)} [({\cal D}\phi)_{i_1\cdots i_n 5}^{(0)}, \Lam_{i_{n+1}}^{(0)}],
\ee
where the exterior derivative of $\phi_{i_1 \cdots i_{n-1} 5}$ is defined by
\be
({\cal D}\phi)_{i_1\cdots i_n 5}^{(0)}
= g\sum_{(n)} [D_{i_1}, \phi_{i_2\cdots i_n 5}^{(0)}].
\ee
It then can be possible to down covariant equations of motion 
using exterior derivatives and totally antisymmetrized tensors.

\subsection{Non-Abelianizing the Abelian Self-Dual Theory}

We have mentioned that the anomalous transformation of $H^{(0)}_{ijk}$ \eqref{w} will cause the problem to define a gauge invariant action. For example, if we define a Yang-Mills-like theory, the Lagrangian should look like
\be
\frac{1}{6}\mbox{Tr}(H_{ijk}^{(0)}H^{(0)}{}^{ijk} + 3H_{ij5}^{(0)}H^{(0)}{}^{ij5}
+ H_{ijk}^{(KK)}H^{(KK)}{}^{ijk} + 3H_{ij5}^{(KK)}H^{(KK)}{}^{ij5}).
\ee
Only the first term is not gauge invariant.
It is not clear how to modify the action to make it invariant.
Similarly it is hard to define the usual kinetic term for 
the components $\phi_{i_1\cdots i_n}^{(0)}$ of a matter field. In the following we will see that in a Lagrangian formulation
of the non-Abelian self-dual gauge theory in 6 dimensions,
we do not have to use the variables $B_{ij}^{(0)}$ explicitly,
so these anomalous covariant transformation laws will never be used. In fact,  we will simply define $H_{ijk}^{(0)}$ to be the Hodge dual of $F_{ij}$
\bea
H_{ijk}^{(0)}\equiv {1\over 2}\eps_{ijklm} F^{lm} \ ,
\eea
so that its gauge transformation is the same as other components.
As a result the covariant transformation laws for matter fields can be uniformly defined as
\be
\d \Phi = g[\Phi, \Lam_5^{(0)}]
\ee for all components of a matter field.
\\

Recall that the Abelian action for chiral 2-form is \eqref{51} (we can obtain it from a gauge-fixed PST action \eqref{c51} as mentioned before)
\be
S = \frac{1}{4} T_{M5}T_{M2}^{-2} \int d^6 x \; \left(
\frac{1}{6}\eps^{ijklm}H_{ijk}\left[
H_{lm5} + \frac{1}{6}\eps_{lmnpq} H^{npq}
\right]
\right).
\ee If we consider the compactification of the Abelian theory on a circle of radius $R$ along $x^5$, all fields then are decomposed into their zero modes and KK modes and the action becomes
\be
S = S^{(0)} + S^{(KK)},
\ee
where
\bea
S^{(0)} &=& \frac{2\pi R}{12} T_{M5}T_{M2}^{-2} \int d^5 x \; H^{\z}_{ijk}H^{\z}{}^{ijk} \ , \\
S^{(KK)} &=& \frac{1}{4} T_{M5}T_{M2}^{-2} \int d^6 x \; \left(
\frac{1}{6}\eps^{ijklm}H^{(KK)}_{ijk}\left[H^{(KK)}_{lm5} + \frac{1}{6}\eps_{lmnpq} H^{(KK)}{}^{npq}
\right]\right) \ . \nn\\
\eea
The zero modes $B_{ij}^{(0)}$ are 5 dimensional 2-form potential. One can carry out the standard procedure of electric-magnetic duality
for $S^{(0)}$ to get an action for the dual 1-form potential
\be
S^{\z}_{dual} = \frac{2\pi R}{4} T_{M5}T_{M2}^{-2} \int d^5 x \; F_{ij} F^{ij}\ ,
\ee
where $F_{ij} = H^{(0)}_{ij5}$ is the field strength of the dual 1-form potential $B_{i5}^{(0)}$.
\\

Let us see that how the equations of motion derived from the action
$S^{(0)}_{dual} + S^{(KK)}$ leads to configurations satisfying self-duality conditions. For the zero modes, the equation of motion derived from the action $S^{\z}_{dual}$ is
\be
\del^j F_{ij} = 0.
\ee
Defining a 3-form field $H$ by
\be
H_{ijk}^{(0)} = \frac{1}{2} \eps_{ijklm} F^{lm},
\label{HF}
\ee
we see that, due to the equation of motion $\del^j F_{ij} = 0$ , a 2-form potential $B^{(0)}$ exists locally
such that $H^{(0)} = dB^{(0)}$. Since $F$ also satisfies the Jacobi identity $dF = 0$,
we also have
\be
\del^k H_{ijk}^{(0)} = 0.
\ee
Note that our definitioan of $H_{ijk}^{(0)}$ is identical to the self-duality condition
for the zero modes
\be
H_{ijk}^{(0)} = \frac{1}{2}\eps_{ijklm} H^{(0)}{}^{lm5} \ .
\ee
Hence we see that the zero modes of the self-dual gauge field
can be simply described by the 5D Maxwell action.
\\

It is natural to non-Abelianize the equation of motion for the zero modes by
\be
[D^j, F_{ij}] = 0 + \cdots \label{1}
\ee
up to additional covariant terms that vanish when the Lie algebra $G$ is Abelian.
In the next section we will derive the complete equation from an action principle. Let us stress again that for the non-Abelian theory,
we still define $H^{(0)}_{ijk}$ simply as the Hodge dual of $F_{ij}$,
hence it is not necessary to introduce the new definition of  $H^{(0)}_{ijk}$ which has the unusual transformation law.
The transformation of $F_{ij}$ would then imply that $H^{(0)}$, transforms simply as
\be
\d H^{(0)}_{ijk} = [H^{(0)}_{ijk}, \Lambda^{(0)}_5].
\ee 

On the other hand, for the KK modes, the equations of motion derived from varying $S^{(KK)}$ is
\be
\eps^{ijklm} \del_k \left(H^{(KK)}_{lm5} + \frac{1}{6}\eps_{lmnpq} H^{(KK) npq}\right)
= 0 \ .
\ee
This implies that
\be
\eps^{ijklm}\left(H^{(KK)}_{lm5} + \frac{1}{6}\eps_{lmnpq} H^{(KK) npq}\right)
= \eps^{ijklm}\Phi^{(KK)}_{lm} \ ,
\ee
for some tensor $\Phi_{lm}^{(KK)}$ satisfying
\be
\eps^{ijklm}\del_k \Phi^{(KK)}_{lm} = 0.
\ee
We can redefine $B_{lm}^{(KK)}$ by a shift
\be
B_{lm}^{(KK)} \rightarrow B_{lm}^{' (KK)} \equiv B_{lm}^{(KK)} + \del_5^{-1} \Phi_{lm}^{(KK)} \ ,
\ee
such that
\bea
H^{(KK)}_{lm5} &\rightarrow& H^{' (KK)}_{lm5} \equiv H^{(KK)}_{lm5} + \Phi^{(KK)}_{lm}, \\
\eps^{lmnpq} H^{(KK)}_{npq} &\rightarrow& \eps^{lmnpq} H^{' (KK)}_{npq}= \eps^{lmnpq} H^{(KK)}_{npq} \ .
\eea
As a result of this shift (This shift is also a gauge symmetry of this theory), we have the self-duality condition
\be
H^{(KK)}_{lm5} = -\frac{1}{6}\eps_{lmnpq} H^{(KK) npq}.
\ee
\\

Let us now consider the non-Abelian counterpart of the equation of motion of KK modes as
\be
\eps^{ijklm} \left[D_k, \left(H^{(KK)}_{lm5}
+ \frac{1}{6} \eps_{lmnpq} H^{(KK) npq}\right)\right]
= 0.\label{2}
\ee
This implies that
\be
\eps^{ijklm} \left(H^{(KK)}_{lm5} + \frac{1}{6} \eps_{lmnpq} H^{(KK) npq}\right)
= \eps^{ijklm} \Phi^{(KK)}_{lm},
\ee
where $\Phi^{(KK)}_{lm}$ satisfies
\be
\eps^{ijklm} [D_k, \Phi^{(KK)}_{lm}] = 0.
\ee
This again can be absorbed into a shift of $B_{lm}^{(KK)}$
\be
B_{lm}^{(KK)} \rightarrow B_{lm}^{' (KK)} \equiv B_{lm}^{(KK)} + \del_5^{-1} \Phi_{lm}^{(KK)},
\ee
so that the self-duality condition is arrived. Notice that it is the KK modes that allows us to consider this ''non-local'' shift as this non-local operator only consistenly acts on KK modes. Also note that the gauge transformation parameter $\Phi^{(KK)}_{lm}$ has to transform covariantly 
\be
\d \Phi^{(KK)}_{lm} = [\Phi^{(KK)}_{lm}, \Lam_5^{(0)}],
\ee
because the constraint of it is covariant. It can then be checked that this extra shift commutes with 
the gauge transformation.  
\\

Our task in the next section is to give an action that would lead to
the non-Abelian equations of motion \eqref{1}, \eqref{2} , which we have shown here that it will give the self-duality condition of 2-form potentials.

\subsection{Action}

We consider the following action for the non-Abelian chiral 2-form potential (the gauge sector of multiple M5-branes on $S^1$)
\be
S = S^{(0)} + S^{(KK)},
\ee
where
\bea
S^{(0)} &=& \frac{2\pi R}{4} T_{M5}T_{M2}^{-2} \int d^5 x \; \mbox{Tr}(F_{ij}F^{ij}),
\label{S0-1} \\
S^{(KK)} &=& \frac{1}{4} T_{M5}T_{M2}^{-2} \int d^6 x \; \mbox{Tr}\left(
\frac{1}{6} \eps^{ijklm}H_{ijk}^{\nz}\left[
H_{lm5}^{(KK)} + \frac{1}{6}\eps_{lmnpq} H^{(KK)}{}^{npq}
\right]
\right) \ .\nn\\
\eea
This gauge invariant action
is a straightforward generalization of the action
for the Abelian theory. For small $R$,
the M5-branes should be approximated by D4-branes in the type $\IIA$ string theory,
so $S^{(0)}$ should be identified as the Yang-Mills theory for multiple D4-branes
\be
S^{(0)} = \frac{1}{4} T_{D4} T_s^{-2} \int d^5 x \; \mbox{Tr}(f_{ij}f^{ij}),
\ee
where the field strength $f_{ij}$ for multiple D4-branes is
\be
f_{ij} \equiv [\del_i + A_i, \del_j + A_j] = \del_i A_j - \del_j A_i + [A_i, A_j].
\ee
It is known that the gauge potential $A$ in D4-brane theory
is related to the gauge potential $B$ in M5-brane theory via the relation
\be
A_i = 2\pi R B_{i5}^{(0)}.
\ee
Plugging in the values of the parameters involved,
\be
T_{M5} = \frac{1}{2\pi} T_{M2}^2, \qquad
T_{D4} = \frac{1}{(2\pi)^4 g_s \ell_s^5}, \qquad
T_s = \frac{1}{2\pi \ell_s^2}, \qquad
R = g_s \ell_s,
\ee
we find that the coupling constant should be given by
\be
g = 2\pi R.
\ee
This factor can also be obtained by demanding that
the soliton solutions which resemble instantons in the spatial 4 dimensions
have momentum equal to $n/R$ for some integer $n$ in the $x^5$ direction.
\\

Notice that the overall factor of $2\pi R$ due to the integration over $x^5$
will be multiplied by a factor of $1/g^2$ where $g$ is the Yang-Mills coupling for the zero mode field strength $F_{ij}$,
giving an overall factor of $1/R$,
in agreement with the requirement of conformal symmetry in 6 dimensions \cite{Witten}. Normally the coupling constant of an interacting field theory is
independent of whether the space is compactified.
Our strategy is to define a 6 dimensional field theory as
the decompactification limit of a compactified theory,
and the coupling depends on the compactification radius.
In some sense, the coupling constant $g$
is not really the coupling of the decompactified theory,
which is a conformal field theory without any free parameter.
\\

Assuming that we will be able to show in future works that
a well defined theory does exist in uncompactified 6 dimensional spacetime
as the decompactification limit of our model,
one would still wonder how such a theory
can be fully Lorentz invariant,
while its definition involves the choice of a special direction.
We will discuss more on this point later.
\\

The variation of $S^{(0)}$ leads to Yang-Mills equations,
which can be interpreted as the self-dual equation for the zero modes.
The full equation of motion for the zero modes $B_{i5}^{(0)}$
should also include variations of $S^{(KK)}$,
which modifies the Yang-Mills equation by commutators
that vanish in the Abelian case. Explicitly, the equations of motion is
\bea
[D_j, F^{ij}] = \frac{1}{2} \int_0^{2\pi R} dx^5 \;
\left[\hat{B}_{jk}, \left(H^{(KK)}{}^{ijk} - \frac{1}{4} \eps^{ijklm} H^{(KK)}_{lm5}\right)\right] \ .
\eea

A useful feature of $S^{(KK)}$ is that it depends on
$B_{i5}^{(KK)}$ and $B^{(KK)}_{ij}$ only through $\hat{B}_{ij}^{(KK)}$.
Therefore, we only need to consider the variation of $\hat{B}_{ij}^{(KK)}$. Explicitly, the equations of motion is
\bea
\eps^{ijklm} \left[D_k, \left(H^{(KK)}_{lm5}
+ \frac{1}{6} \eps_{lmnpq} H^{(KK) npq}\right)\right] = 0
\eea which leads to the equation of motion that is equivalent to the self-duality condition 
via a shift in $B_{lm}^{(KK)}$ as we explained in the previous section.
\\

Notice that despite the appearance of the nonlocal operator $\del_5^{-1}$
and the nonlocal separation of KK modes from zero modes 
in the 5-th direction, 
this action  
is an ordinary local action from the viewpoint of the uncompactified 5 dimensional Minkowski space:
\bea
S &=& 
\frac{2\pi R}{4} T_{M5}T_{M2}^{-2} \int d^5 x \; \mbox{Tr} \Big\{
F_{ij}F^{ij}
\nn \\
&+&
\frac{1}{6} \sum_{p\in\mathbb{Z}}
\eps^{ijklm}h_{ijk}(-p)\left[
\left(i\frac{p}{R}\right)\hat{b}_{lm}(p) 
+ \frac{1}{6}\eps_{lmnpq} h^{npq}(p)
\right]
\Big\},
\eea
where $\hat{b}_{ij}(p)$ and $h_{ijk}(p)$ are the KK mode coefficients of
$\hat{B}_{ij}$ and $H_{ijk}$ defined by
\bea
\hat{B}_{ij}(\vec{x}, x^5) &=& \sum_{p\in\mathbb{Z}} \hat{b}_{ij}(\vec{x}, p) e^{ipx^5/R}, \\
H_{ijk}(\vec{x}, x^5) &=& \sum_{(i, j, k)} \sum_{p\in\mathbb{Z}} 
\left(\del_i \hat{b}_{jk}(\vec{x}, p) 
+ \sum_{q\in\mathbb{Z}} [A_i(\vec{x}, q), \hat{b}_{jk}(\vec{x}, p-q)]\right) e^{ipx^5/R} \ , \nn\\
\eea
where $\vec{x} = (x^0, x^1, \cdots, x^4)$.
The derivation of the equations of motion above can be viewed 
as a collective expression of equations of motion derived by 
varying each KK mode or zero mode one at a time.  It is clear from these expressions that 
our theory is local in the directions of $\vec{x}$, 
and its nonlocality is restricted to the $x^5$-direction. At this moment we can not prove or disprove the causality 
in the direction of $x^5$. At least causality is apparently still preserved in the uncompactified directions $\vec{x}$.

\subsection{Comments}

In this formulation, we have avoided commutators
involving two KK modes, e.g. terms of the form $[B^{(KK)}, \Lambda^{(KK)}]$. Correspondingly, there is no term of the form $[B^{(KK)}, B^{(KK)}]$ in the equations of motion or action. In fact, all gauge interactions are mediated via zero modes. Here is our interpretation.
In the limit $R \rightarrow \infty$,
the Fourier expansion of a field approaches to the Fourier transform
\be
\Phi(x^5) = \sum_n \Phi_n e^{in x^5/R}
\qquad \longrightarrow \qquad
\Phi(x^5) = \int \frac{dk_5}{2\pi} \; \tilde{\Phi}(k_5) e^{ik_5 x^5}.
\ee
The coefficients $\Phi_n$ approach to $\tilde{\Phi}(k_5)$ as
\be
\tilde{\Phi}(k_5) = 2\pi R \Phi_n \qquad
(k_5 = n/R).
\label{PhiPhi}
\ee
According to this expression,
the value of a specific Fourier mode $\Phi_n$ must approach to zero
in the limit $R \rightarrow \infty$.
In particular, the amplitude of the zero mode approaches to zero.
While all interactions are mediated via the zero mode,
this does not imply that there is no interaction in the infinite $R$ limit,
because the coupling $g = 2\pi R \rightarrow \infty$.
The product of the amplitude of the zero mode with the coupling
is actually kept finite in the limit. 
\\

In the limit $R \rightarrow \infty$,
the KK modes $B_{\mu\nu}^{(KK)}$ should be identified with the 2-form potential
in uncompactified 6 dimensional spacetime.
In uncompactified space,
the constant part of $B_{\mu\nu}$ is not an observable,
hence physically the KK modes $B_{\mu\nu}^{\nz}$ do not
miss any physical information a 2-form potential can carry.
\\

Regarding the large R limit, we consider that the zero modes $B_{\mu\nu}^{\z}$ approach to zero but
a new field $A_i$ replacing $2\pi RB_{i5}^{\z}$ survives the large $R$ limit.
The field $A_i$ can not be viewed as part of the 2-form potential,
in the sense that, due to the infinite scaling of $B_{i5}^{(0)}$ by $R$,
it can not be combined with $B_{\mu\nu}^{(KK)}$
in a Lorentz covariant way to form a new tensor in 6 dimensions.
Rather it should be understood as the 1-form needed
to define gerbes (or some similar geometrical structure)
together with the 2-form potential.
However this does not increase the physical degrees of freedom of the 6 dimensional theory
in the sense that the number of physical degrees of freedom in the 5 dimensional field $A_i$
is negligible compared with that of a 6 dimensional field.
\\

The fact that gauge transformation laws do not have
terms of the form $[B^{(KK)}, \Lambda^{(KK)}]$,
and the fact that the equations of motion
do not have terms of the form $[B^{(KK)}, B^{(KK)}]$,
are both telling us that our model is linearized with respect to the 2-form potential.
No self-interaction of the 2-form potential is present,
and all interactions are mediated by the 1-form potential $A_i$. As the decompactification limit $R \rightarrow \infty$
is also the strong coupling limit $g \rightarrow \infty$,
we do not expect that the classical equations of motion
could give a good approximation for the quantum theory. 
\\

On the other hand, the interpretation above allows us to understand some puzzles
about the proposal of recent papers \cite{Douglas,LPS} claiming that
the 5 dimensional D4-brane theory is already sufficient
to describe the 6 dimensional M5-brane system even for a finite $R$.
In their proposal, the momentum $p_5$ in the 5-th (compactified) direction is
represented by the ``instanton'' number on the 4 spatial dimensions. The first puzzle with this interpretaion is that,
in the phase when $U(N)$ symmetry is broken to $U(1)^{N}$,
there is no instanton solution.
But physically this corresponds to having M5-brane well separated
from each other,
and they should still be allowed to have nonzero $p_5$.
This problem does not exist in our model.
In our model $p_5$ is carried by the KK modes
when the Lie algebra of the gauge symmetry is Abelian.
Furthermore, the Abelian case of our model is already known to
be equivalent to a 6 dimensional theory which has
the full Lorentz symmetry in the large $R$ limit.
\\

The second puzzle of their proposal is that
the instanton number only gives the total value of $p_5$ of a state,
and there is no way to specify the distribution of $p_5$ over
different physical degrees of freedom.
For example, the state with $m$ units of $p_5$ contributed from
the scalar field $X^1$ and $n$ units of $p_5$ from $X^2$
cannot be distinguished from the state with the numbers $m$ and $n$ switched.
A possible resolution of this puzzle is that,
perhaps due to strong interactions, 
we can no longer distinguish different distributions of $p_5$ 
over different degrees of freedom. 
In other words,
it is unphysical to specify the distribution of $p_5$.
If true, this would be a mysterious phenomenon,
but we can not rule out this possibility,
as we know very little about how to label physical degrees of freedom 
in a strongly correlated system. In our model, the instanton number of the 1-form
$ A_i \equiv RB_{i5}^{\z} $
should only be interpreted as the value of $p_5$ of the field $A_i$.
In other words, the so-called ``zero-modes'' $B_{i5}^{(0)}$ can still carry nonzero $p_5$.
The 5-th momentum of the 2-form potential is manifest as the KK mode index.
The scalar fields $X^I$ and the fermions $\Psi$,
when they are introduced into our model,
would have their own KK modes to specify their $p_5$ contribution.
There is no ambiguity in the momentum carrier for a given instanton number.
\\

The reader may wonder whether it is redundant or over-counting
for $A_i$ to be able to carry nontrivial $p_5$.
After all, $A_i$ is just $B_{i5}^{(0)}$ rescaled.
Has not the KK modes $B_{i5}^{(KK)}$ already taken care of
the contribution of $B_{i5}$ to $p_5$?
How can a field carry momentum in the $x^5$-direction
if it has no fluctuation (e.g. propagating wave) in that direction?
The answer is simple.
It is well known in classical electrocmagnetism that
the simultaneous presence of constant electric and magnetic fields
carry momentum,
because the momentum density $p_i$ is proportional to $F^{0j}F_{ij}$.
In the temporal gauge
, $A_0 = 0$,
the conjugate momentum of $A_j$ is 
$ \Pi^j \equiv \del_0 A^j $
and the momentum density $p_i$ is proportional to
\be
F^{0j}F_{ij} = \Pi^j (\del_i A_j) - \Pi^j (\del_j A_i).
\ee
The first term is the standard contribution of a field to momentum $p_i$.
We also have $(\del_0 \phi) (\del_i \phi)$ for a scalar field $\phi$.
But there is no analogue of the 2nd term for a scalar field.
It is possible for the 2nd term to be present because $A_i$ has a Lorentz index.
The zero mode of $A_i$ in the $x^i$ direction can also contribute to $p_i$
through this term. Similarly, for a 3-form field strength $H$,
the momentum density of $p_5$ is proportional to $H_{0ab}H^{ab5}$
($a, b = 1, 2, 3, 4$),
which includes the zero mode contribution
\be
H_{0ab}^{\z}F^{ab} = \frac{1}{6}\eps_{0abcd5} F^{ab}F^{cd} \ ,
\ee
because $H^{\z}_{ab5} = F_{ab}$.
This is precisely the same expression as the instanton number density.
Note that there are also contributions to $p_5$ from the KK modes
$H_{0ab}^{(KK)}H^{(KK)}{}^{ab5}$ in addition to the zero mode contribution.

\newpage
\section{BRST-Antifields Quantization on Non-Abelian 2-Form}
\subsection{BRST Symmetry of Non-Abelian 2-form}
The quantization of self-dual fields is a long-standing subtle problem. 
One approach toward a quantized M5-brane (self-dual fields) was  developed in \cite {WW}. It is suggested that one can start from a non-chiral action, which has a well-defined partition function denoted by $Z$, then consider to write $Z$ as an absolute value squared of the chiral field's partition function.  So as  an intermediate step toward the quantization of M5-branes, it might be useful to firstly consider Z, which corresponds to the partition function of non-Abelian 2-form potential $without$ imposing the self-duality condition. In this section, although the Lorentz covariance is still broken, we investigate the BRST transformation laws of this non-local non-Abelian 2-form gauge theory.  We will also give a BRST invaraint gauge-fixed action.
\\

The quantization of reducible gauge theories is not that straightforward. As we know, a gauge-fixing procedure is needed to render dynamical degrees of freedom by using ghost fields, which are used to compensate for the gauge degrees of freedom. In reducible gauge theories, some so-called ''ghosts of ghosts'' will be needed and the ordinary Faddeev-Popov procedure become quite complicated for this kind of reducible theories. Here we will use BRST-antifield formulation (or Batalin-Vilkovisky method \cite{BV}) to deal with this situation.
\\

The Yang-Mills-like action for the non-Abelian 2-form is given by
\be
S = S^{(0)} + S^{(KK)},
\ee
where
\bea
S^{(0)}&=&{2\pi R \over 4}\int d^5x~ Tr \Big[ F^{(0)ij}F^{(0)}_{ij}\Big]\ , \\
S^{(KK)}&=&\int d^6x~ Tr \Big[{1\over 6} H^{(KK)ijk}H^{(KK)}_{ijk}+ {1\over 2} H^{(KK)ij5}H^{(KK)}_{ij5}\Big] \ ,
\eea 
where we separately define zero modes and KK modes as we did before. Note this is just the standard form of an action that is different from the self-dual action we have discussed so far.  At first we will need to find the existence of the BRST transformation laws of this theory where gauge parameters such as \bea \Lam^{(0)}_5, \Lam^{(KK)}_5,\Lam^{(KK)}_i \eea become ''ghost fields'' with fermionic statistics and join BRST transformation laws. In the following we use the same notation for the gauge parameters but we should re-interpret gauge parameters as ghost fields. 
\\

The BRST transformation laws denoted by the BRST operator $s$ for the non-Abelian 2-form theory are given by
\bea
s B^{(0)}_{i5}&=&[D_i, \Lam^{(0)}_{5}] \ ,\\
s B^{(KK)}_{i5}&=&[D_i,\Lam^{(KK)}_5]-\del_5\Lam^{(KK)}_i+g[B^{(KK)}_{i5},\Lam^{(0)}_5]\ ,\\
s B^{(KK)}_{ij}&=&[D_i,\Lam^{(KK)}_j]-[D_j\Lam^{(KK)}_i]+g[B_{ij},\Lam^{(0)}_5]\nn\\
&&-g[F_{ij},\del^{-1}_5 \Lam^{(KK)}_5]\ ,\\
s \Lam^{(0)}_5&=&-{g\over2} [\Lam^{(0)}_5,\Lam^{(0)}_5]\ ,\\
s \Lam^{(KK)}_5&=&\del_5 \alpha-g[\Lam^{(0)}_5,\Lam^{(KK)}_5]\ ,\\
s \Lam^{(KK)}_i&=&[D_i, \alpha]-g[\Lam^{(0)}_5,\Lam^{(KK)}_i]\ ,\\
s\alpha &=& g[\alpha,\Lam^{(0)}_5]\ ,
\eea
where $\alpha$ is a commuting ghost representing the redundancy in the theory. Notice that $\alpha$ appears only as a KK mode but here we ignore the KK index for simplicity.  We again will not use $\Lam^{(0)}_i$ explicitly just like we did not use $B^{(0)}_i$ explicity in our theory. The BRST operator $s$ has ghost number one and the ghost of ghost ''$\alpha$'' has ghost number two which can be read from the BRST transformation laws. The role of $\alpha$ is to fix the residual degrees of freedom coming from the gauge symmetry of gauge symmetry of the theory as mentioned above. These BRST transformation laws satisfy 
\bea
s^2=0 \ ,
\eea which is the nilpotency condition for the BRST transformation. The existence of the BRST symmetry is crucial for further analysis on quantizaton of a gauge thoery. Physical observables are defined as those BRST invariant (closed) but can not be expressed as a BRST variation of something (exact). In short, observables correspond to the elements of the BRST cohomology.

\subsection{Field-Antifield Quantization on Non-Abelian 2-form}
Now we consider the gauge fixing process following the field-antifield method \cite{BV}, where the original configuration space will be enlarged to include extra fields such as ghost fields and ghosts for ghosts. The minimal action is given by
\bea
S_0=S^{(0)}_0+S^{(KK)}_0 \ ,
\eea
where zero modes' part is given by
\bea
S^{(0)}_0={2\pi R \over 4}\int d^5x~ Tr\Big[ (F^{ij}F_{ij}+B^{*(0)i5}\Big([D_i, \Lam^{(0)}_{5}]\Big)-\Lam^{*(0)}_5\Big({g\over2} [\Lam^{(0)}_5,\Lam^{(0)}_5]\Big)\Big] \ ,
\eea
and KK modes' part reads
\bea
S^{(KK)}_0&=&\int d^6x~Tr~\Big\{ \Big[{1\over 6} H^{ijk}H_{ijk}+ {1\over 2} H^{ij5}H_{ij5}+B^{*(KK) i5} \Big([D_i,\Lam^{(KK)}_5]-\del_5\Lam_i\nn\\
&&+g[B^{(KK)}_{i5},\Lam^{(0)}_5]\Big)
+B^{*(KK) ij} \Big([D_i,\Lam^{(KK)}_j]-[D_j\Lam^{(KK)}_i]+g[B_{ij},\Lam^{(0)}_5]\nn\\
&&-g[F_{ij},\del^{-1}_5 \Lam^{(KK)}_5]\Big)\Big]+\Lam^{*(KK)}_5\Big(\del_5 \alpha-g[\Lam^{(0)}_5,\Lam^{(KK)}_5]\Big)+\Lam^{*(KK)i}\Big([D_i, \alpha]\nn\\
&&-g[\Lam^{(0)}_5,\Lam^{(KK)}_i]\Big)+\alpha^* \Big(g[\alpha,\Lam^{(0)}_5]\Big)\Big\} \ .
\eea 
Here $(B^{*(KK)}_{i5}, B^{*(KK)}_{ij}, B^{*(0)}_{i5}, \Lam^{*(0)}_5, \Lam^{*(KK)}_5, \Lam^{*(KK)}_i, \alpha^* )$ are antifields introduced as the sources of the BRST transformation laws of corresponding gauge fields. The ghost number's relation between a field and its corresponding antifield is given by
\bea
gh(\Phi)+gh(\Phi^*)=-1 \ ,
\eea which can be read from the fact that the action has vanishing ghost number. We also notice that the relation of (mass) dimension between  a field and its corresponding antifield is $D(\Phi)+D(\Phi^*)=6$, thus for instance, $D(B_{\m\n})=2$ so that $D({B^*_{\m\n})}=4$, and the coupling constant has $D(g)=-1$. 
\\

Antifields will be eliminated by using the so-called gauge-fixing fermion $\Psi$ via 
\bea \Phi^{*I}={\del \Psi\over \del \Phi^I} \ .
\eea 
 The partition function constructed by a gauge-fixed effective action can be written as
\bea
Z=\int {\cal D}\Phi^I {\cal D}\Phi^{*I} \delta\Big({\Phi^{*I}-{\del \Psi\over \del \Phi^I}}\Big)~e^{{i\over\hbar} S_{eff}} \ ,
\eea
where $\Phi^I$ represents all fields in the theory. Since $\Psi$ must be a functional of fields only (not the antifields) and its ghost number is $-1$, we see that it is not possible to write down an acceptable gauge fermion unless we introduce extra fields. Thus we introduce the following trivial pairs (doublet) defined by (A,B) with the following relations
\bea
sA=B~~;~~sB=0 \ . 
\eea The transformation of B is simply the consequence of the nilpotency of A. 
\\

The zero modes' part is an irreducible system. It is sufficient to introduce only one doublet.  For the KK modes' part, it is a reducible system and we will introduce three pairs. The corresponding extend actions are given by
\bea
S^{(0)}_{ex}&=&S^{(0)}_0-{2\pi R\over 4}\int d^5x~Tr~ \bar {D^*} E\ , \\
S^{(KK)}_{ex}&=&S^{(KK)}_0-\int d^6x~Tr~ \Big( \bar \Lam^{*i} b_i+\bar\Lam^{*5} b_5+\bar\alpha^{*} b-\bar\omega^*\pi \Big) \ ,
\eea where we introduce doublets
\bea
&&(\bar D, E)~~\text{for~ zero~modes} \ , \\
&&(\bar\Lam^{\m}, b_\m),~ (\bar\alpha, b),~(\bar\omega,\pi)~\text{for~KK~ modes} \ .
\eea  Note that fields in the same doublet have the same (mass) dimension.
\\

As an example, let us consider the following the gauge fermion $\Psi$
\bea
\Psi^{(0)}&=&{2\pi R\over 4}\int d^5x~Tr~ \bar D (- {E\over 2\eta}+\del^i B^{(0)}_{i5}) \ ,\\
\Psi^{(KK)}&=&\int d^6x~Tr~\Big(\bar \Lam^i (\del^j B^{(KK)}_{ij}+\del ^5 B^{(KK)}_{i5}+\del_i\bar\omega)\nn\\
&&+\bar\Lam^5(\del^i B^{(KK)}_{i5}+\del_5 \bar\omega)+\bar\alpha\del^\m\Lam^{(KK)}_\m\Big) \ ,
\eea 
where $\eta$ is an adjustable parameter. By $\Phi^*={\del \Psi\over \del \Phi}$ (Recall that $\Psi$ is a functional of fields only.), we obtain the gauge fixing action:
\bea
S^{(0)}_{gf}&=& {2\pi R \over 4}\int d^5x~ Tr \Big[ (F^{(0)ij}F^{(0)}_{ij})-\del^{i}\bar D\Big([D_i, \Lam^{(0)}_{5}]\Big)+( {E\over 2\eta}-\del^i B^{(0)}_{i5}) E\Big]\\
S^{(KK)}_{gf}&=&\int d^6x~Tr~\Big\{{1\over 6} H^{(KK)ijk}H^{(KK)}_{ijk}+ {1\over 2} H^{(KK)ij5}H_{(KK)ij5}\nn\\
&&-\del^5 \bar\Lam^{(KK)i}\Big([D_i,\Lam^{(KK)}_5]-\del_5\Lam^{(KK)}_i+g[B^{(KK)}_{i5},\Lam^{(0)}_5]\Big)\nn\\
&&-\del^i \bar\Lam^{(KK)5}\Big([D_i,\Lam^{(KK)}_5]-\del_5\Lam^{(KK)}_i+g[B^{(KK)}_{i5},\Lam^{(0)}_5]\Big)\nn\\
&&-\del^i \bar\Lam^{(KK)j} \Big([D_i,\Lam^{(KK)}_j]-[D_j\Lam^{(KK)}_i]+g[B_{ij},\Lam^{(0)}_5]-g[F_{ij},\del^{-1}_5 \Lam^{(KK)}_5]\Big)\nn\\
&&-\del_5 \bar\alpha\Big(\del_5 \alpha-g[\Lam^{(0)}_5,\Lam^{(KK)}_5]\Big)-\del^i \bar\alpha\Big([D_i, \alpha]-g[\Lam^{(0)}_5,\Lam^{(KK)}_i]\Big)\nn\\
&&-(\del^j B^{(KK)}_{ij}+\del ^5 B^{(KK)}_{i5}+\del_i\bar\omega) b^i-(\del^i B^{(KK)}_{i5}+\del_5 \bar\omega)b_5\nn\\
&&-(\del^\m\Lam^{(KK)}_\m) b-(\del^i\bar\Lam_i+\del^5 \bar \Lam_5)\pi \Big\} \ .
\eea Notice that this gauge-fixed action is indeed invaraint under BRST transformation laws
\bea
s(S^{(0)}_{gf})&=&0 \ ,\\
s(S^{(KK)}_{gf})&=&0 \ ,
\eea and there is no other gauge symmetry left. After we intergrate auxiliary fields out, the guage fixing conditions can be read in the following covaraint forms
\bea
\del^i B^{(0)}_{i5}&=&{E\over \eta} \ ,\\
\del^\n B^{(KK)}_{\m\n}+\del_\m \bar\omega&=&0\ ,\\
\del^\m\Lam^{(KK)}_\m&=&0\ ,\\
\del^\m\bar\Lam_\m&=&0\ .
\eea 
If we take the parameter $\eta\rightarrow \infty$, the Lorenz gauge will be imposed as a delta-function condition for zero modes. The Lorentz covariance is still broken for the KK modes part.  The gauge fixing process here is similiar with the case when one deals with the Abelian 2-form theory.  
\\

Regarding the counting of the degrees of freedom, the ghost fields' degrees of freedom subtract since they represent these unphysical parts. The two-form potentials give $+15$, while ghost fields $\Lam_\m$ and $\bar \Lam_\m$ both give $-6$, then $\omega$, $\alpha$ and $\bar\alpha$ all give $+1$. In total we have 15-6-6+1+1+1=6 physical degrees of freedom, and that is the expected answer for a 2-form potential theory in 6D. 
\\

Intergrating all auxiliary fields out in $S_{gf}$ and denote $S_{eff}$ as the effective action. The partition fuction for this non-abelian 2-form theory can be formally given by 
\bea
Z^{(0)}&=&\int {{\cal D} B^{(0)}_{i5}} {{\cal D} \Lam^{(0)}_{5}} {{\cal D} \bar D}~\delta (\del^i B^{(0)}_{i5}-{E\over \eta})~ e^{{i\over \hbar}S^{(0)}_{eff}}\\
Z^{(KK)}&=&\int {{\cal D} B^{(KK)}_{\m\n}}  {{\cal D} \Lam^{(KK)}_{\m}}{{\cal D} \bar\Lam^{(KK)}_{\m}}{{\cal D} \bar\omega}{{\cal D} \alpha}{{\cal D} \bar\alpha}\nn\\
&&\delta (\del^\n B^{(KK)}_{\m\n}+\del_\m \bar\omega)~\delta(\del^\m\Lam^{(KK)}_\m)~\delta(\del^\m\bar\Lam_\m)~e^{{i\over\hbar} S^{(KK)}_{eff}} \ ,
\eea where the field $\bar D$ in zero modes' part is referred as the anti-ghost field in the Yang-Mills theory. 
\\

One may wonder whether or not it is possible to start from a standard Yang-Mills-like action as the classical action to obtian the self-duality condition at the partition function level via enlarging the configuration space through field-antifield formulation. The final partition function should look like (for KK modes)
\bea
Z^{(KK)}&=&\int {{\cal D} B^{(KK)}_{\m\n}}  {{\cal D} \Lam^{(KK)}_{\m}}{{\cal D} \bar\Lam^{(KK)}_{\m}}{{\cal D} \bar\omega}{{\cal D} \alpha}{{\cal D} \bar\alpha}\nn\\
&&\delta (\del^\n B^{(KK)}_{\m\n}+\del_\m \bar\omega)~\delta(\del^\m\Lam^{(KK)}_\m)~\delta(\del^\m\bar\Lam_\m)~\delta ({H-*H})_{ij5} ~e^{{i\over\hbar}S^{(KK)}_{eff}} \ ,\nn\\
\eea where ${(H-*H})_{ij5}$ represents the self-duality condition. Following the method of field-antifield formulation, the corresponding extend action is 
\bea
S'=\int d^6 x~Tr~P^{*ij}Q_{ij} \ ,
\eea where $(P_{ij}, Q_{ij})$ form a triviar pair. The corresponding extra gauge fermion is
\bea
\Psi'=\int d^6x~ Tr~(H-*H)_{ij5}P^{ij} \ .
\eea We see that in the end althrough we will have a delta function that gives the self-dualty condition after intergrating out $Q_{ij}$. However, the result in not correct since it introduces extra degrees of freedom that come from the kinetic terms of $P_{ij}$. The result indicates that this naive quantization approach for a self-dual theory does not work.

\newpage
\section{A Generalization: Non-Abelian 3-form}

In this section we construct
non-Abelian gauge theory for a 3-form potential. 
Readers who want to focus on the theory of M5-branes might simply skip this section.
\\

We study the Abelian gauge theory for a 3-form potential
on the spacetime of \bea \mathbf{R}^d\times T^2\ .\eea
Let the torus $T^2$ extend in the directions of $x^1$ and $x^2$.
We decompose a field $\Phi$ as 
\be
\Phi = \Phi^{\z} + \Phi^{\nz}\ ,
\ee
where the zero mode $\Phi^{\z}$ has no dependence on $T^2$
\be
\del_a \Phi^{\z} = 0\ .
\ee
The KK mode $\Phi^{\nz}$ can be obtained from $\Phi$ as 
\be
\Phi^{\nz} = \Box^{-1}\Box\Phi\ ,
\ee
where 
\be
\Box \equiv \del^a \del_a 
\qquad
(a = 1, 2).
\ee

The Abelian gauge transformations of a 3-form potential $B$ are given by
\bea
\d B_{i12} &=& \del_i \Lam_{12} - \del_1 \Lam_{i2} + \del_2 \Lam_{i1}, \\
\d B_{ija} &=& \del_i \Lam_{ja} - \del_j \Lam_{ia} + \del_a \Lam_{ij}, \\
\d B_{ijk} &=& \del_i \Lam_{jk} + \del_j \Lam_{ki} + \del_k \Lam_{ij},
\eea
where $a = 1, 2$ and $i, j, k = 0, 3, 4, \cdots, (d+1)$.
There is redundancy in the gauge transformation parameters $\Lam_{ia}, \Lam_{ij}$. The gauge transformation laws are invariant under the transformation
\bea
\d \Lam_{12} &=& \del_1 \lam_2 - \del_2 \lam_1, \\
\d \Lam_{ia} &=& \del_i \lam_a - \del_a \lam_i, \\
\d \Lam_{ij} &=& \del_i \lam_j - \del_j \lam_i.
\eea
Apparently there is also a redundancy in using $\lam$ to parametrize
the redundancy in $\Lam$.
There are $(d+2)$ components in $\lam$,
but only $(d+1)$ of them are independent.
Using the redundancy of $\Lam$,
we can ``gauge away'' $(d+1)$ of the gauge transformation parameters.
For instance, we can set 
\be
\rho_i \equiv \del^a \Lam_{ia} = 0, 
\qquad
\Lam_{12} = 0,
\ee
and use the following gauge transformation parameters
\be
\xi_i \equiv \eps^{ab}\del_a \Lam_{ib}, 
\qquad
\Lam_{ij},
\ee
so that 
\be
\Lam_{ia}^{\nz} = -\eps^{ab}\Box^{-1}\del_b\xi_i,
\ee
and the gauge transformation laws become
\bea
\d B_{i12} &=& -\xi_i, 
\label{dBi12-0} \\
\d B_{ija} &=& -\eps^{ab}\Box^{-1}\del_b(\del_i\xi_j - \del_j\xi_i) + \del_a \Lam_{ij}, 
\label{dBija-0} \\
\d B_{ijk} &=& \del_i \Lam_{jk} + \del_j \Lam_{ki} + \del_k \Lam_{ij}.
\label{dBijk-0}
\eea

Viewing $\Lam_{ia}$ as $d$ copies of 1-form potentials on $T^2$,
the $\xi_i$'s are the corresponding field strengths, 
and so their integrals over $T^2$ are quantized.
It implies that $\xi_i^{\z}$ is quantized,
and so we have to set
\be
\xi_i^{\z} = 0\ ,
\ee
when we use $\xi_i$ as infinitesimal gauge transformation parameters.
In the following, we have $\xi_i = \xi_i^{\nz}$.
\\

Since the original gauge transformation laws have the redundancy,
one can simply carry out the replacement
\bea
\xi_i &\rightarrow& \xi_i - \del_i \Lam_{12}, \\
\Lam_{ij} &\rightarrow& \Lam_{ij} + \Box^{-1}(\del_i\rho_j - \del_j\rho_i).
\eea On the torus $T^2$,
the gauge transformation parameter
$\Lam_{12}^{\z}$ corresponds to a Wilson surface degree of freedom
for the 2-form $\Lam$.
\\

To construct a consistent non-Abelian gauge transformation algebra for the 3-form potential,
we only need to consider transformation laws for the parameters $\xi_i^{\nz}$, $\Lam_{ij}$ and $\Lam_{12}^{\z}$.
In the end we get the full gauge transformation laws through the replacement
\bea
\xi_i^{\nz} &\rightarrow& \xi_i^{\nz} - [D_i, \Lam_{12}^{\nz}], \\
\Lam_{ij}^{\nz} &\rightarrow& \Lam_{ij}^{\nz} 
+ \Box^{-1}([D_i, \rho_j^{\nz}] - [D_j, \rho_i^{\nz}]),
\eea
where the covariant derivative $D_i$ should be defined as
\be
D_i = \del_i + B_{i12}^{\z},
\ee
and
\be
\xi_i^{\nz} \equiv \eps^{ab}\del_a\Lam_{ib}^{\nz}, 
\qquad
\rho_i^{\nz} \equiv \del^a\Lam_{ia}^{\nz}.
\ee
Here we have scaled $B_{i12}$ to absorb the coupling constant $g$,
which is expected to be given by the area $(2\pi)^2 R_1 R_2$ of the torus.
\\

We define the non-Abelian gauge transformations as
\bea
\d B_{i12} &=& [D_i, \Lam_{12}^{\z}] - \xi_i + [B_{i12}^{\nz}, \Lam_{12}^{\z}], \\
\d B_{ija} &=& - \eps^{ab}\Box^{-1}\del_b\left([D_i, \xi_j]-[D_j, \xi_i]\right)
+ \del_a \Lam_{ij} \nn \\
&& + [D_i, \Lam_{ja}^{\z}] - [D_j, \Lam_{ia}^{\z}]
+ [B_{ija}. \Lam_{12}^{\z}], \\
\d B_{ijk} &=& [D_i, \Lam_{jk}] + [D_j, \Lam_{ki}] + [D_k, \Lam_{ij}]
+ [B_{ijk}, \Lam_{12}^{\z}].
\eea
The algebra of gauge transformations is closed
\be
[\d, \d'] = \d'',
\ee
with the parameters of $\d''$ given by
\bea
\Lam_{12}^{\z}{}'' &=& [\Lam_{12}^{\z}, \Lam_{12}^{\z}{}'], \\
\xi_i'' &=& [\xi_i, \Lam_{12}^{\z}{}'] - [\xi_i', \Lam_{12}^{\z}], \\
\Lam_{ij}'' &=& [\Lam_{ij}, \Lam_{12}^{\z}{}'] - [\Lam_{ij}', \Lam_{12}^{\z}].
\eea

The field strengths should be defined as
\bea
H_{ij12}^{\z} &=& [D_i, D_j], \\
H_{ij12}^{\nz} &=& [D_i, B_{j12}^{\nz}]-[D_j, B_{i12}^{\nz}]+\del_4 B_{ij5}^{\nz}-\del_5 B_{ij4}^{\nz}, \\
H_{ijka} &=& [D_i, B_{jka}] + [D_j, B_{kia}] + [D_k, B_{ija}] - \del_a B_{ijk} \nn \\
&& - \eps^{ab} \Box^{-1}\del_b\left( [F_{ij}, B_{k12}^{\nz}] + [F_{jk}, B_{i12}^{\nz}] + [F_{ki}, B_{j12}^{\nz}] \right), \\
H_{ijkl} &=& \sum_{(4)} [D_i, B_{jkl}] - \sum_{(6)} [F_{ij}, \beta_{kl}],
\eea
where
\be
\beta_{ij} \equiv \Box^{-1}\del^a B_{ija},
\ee
so that all the field strength components transform as
\bea
\d H_{ij12} &=& [H_{ij12}, \Lam_{12}^{\z}], \\
\d H_{ijka} &=& [H_{ijka}, \Lam_{12}^{\z}] + \sum_{(3)}[F_{ij}, \Lam_{ka}^{\z}], \\
\d H_{ijkl} &=& [H_{ijkl}, \Lam_{12}^{\z}] + \sum_{(6)}[F_{ij}, \Lam_{kl}^{\z}].
\eea

It may be possible to define a non-Abelian self-dual gauge theory for 
a 3-form potential in 8 dimensional Euclidean space. 
And the same idea can be used to define 
a non-Abelian gauge symmetry for $p$-form potentials 
on $\mathbf{R}^d \times T^{p-1}$.

\newpage
\section{Toward Non-Abelian PST Action and Conclusion}
\subsection{Non-Abelian PST Action}

One of the most interesting problems is to find a manifest Lorentz-covariant formulation of the non-Abelian self-dual 2-form theory. The covariant theory should reproduce the previous 5+1 non-covariant action under proper gauge-fixing conditions. In short, the question is: Can we have a non-Abelian PST-like formulation for multiple M5-branes?
\\

Let me conclude this thesis by pointing out some main difficulties when one wants to find a manifest Lorentz-covariant formulation of this non-Abelian self-dual 2-form theory when using the similar ideas of introducing auxiliary fields.
\\

We should firstly understand that in the case of this non-Abelian chiral 2-form, the non-covariance is very complicated. In the Abelian case, the field strength $H=dB$ is defined covariantly.
(This means that its Maxwell-like action is obviously Lorentz covaraint.) We then can start thinking about how to obtain a covariant self-dual (PST) action.
In the non-Abelian case, however, it Yang-Mills-like action ($s= \int H^2$) is not Lorentz covariant: We treat the 5-th direction differently even before we consider its self-dual formulation.
\\

One can check that the transformed
field equation of the Yang-Mills-like action of this non-abelian 2-form theory under the standard Lorentz transformation is $not$ up to the field equation itself. One also observes that $A_5$ is absent in the gauge algebra,
which is like a gauge choice that violates the Lorentz covariance. A possibility is that our current theory might still under a certain larger gauge, similar to that the PS model can be viewed as a gauge-fixed PST model. (Similar spirit but not exactly since the non-covariance appears in the original gauge symmetry structure.)  
\\

Assuming we have a covariant version of the non-abelian 2-form theory, which means that we first assume $H_{\m\n\lam}H^{\m\n\lam}$ is Lorentz invariant (we will discuss about it more later), is it possible to have a non-Abelian PST self-dual action? Let us introduce an ''Abelian'' auxiliary field $b(x)$ to construct the covariant self-dual theory. (The reason for ''Abelian'' is that we hope that the field $b(x)$ can be fixed to $x^5$ in the end, and $x^5$ should not take a matrix value.). Let us directly non-abelianize the 6D PST action:
\bea
\label{npst}
S={1\over 4} \int d^6x~Tr \Big ( -{1\over 3!} H^{\m\n\lambda}H_{\m\n\lambda}+{1\over 2} {\cal H^{\m\n\rho}} P^\sigma_\rho {\cal H_{\m\n\sigma}}\Big) \ .
\eea 
The variation of the action\footnote{Here we consider $H_{\m\n\lam}=[D_\m, \hat B_{\n\lam}]+[D_\n, \hat B_{\lam\m}]+[D_\lam, \hat B_{\m\n}]$ for a $covaraint$ (if exists) non-abelian 2-form theory under a gauge fixing condition so that we adopt the variable $\hat B$. We denote the connection as $A_\m$ (if exists in 6D). We will detail these issues more later.} gives
\bea
\delta S=\int d^6 x~Tr~\Big( \eps^{\alpha\beta\rho\m\n\lambda}\partial_\m b [D_\rho ,\bar H_{\n\lambda} ] (\delta \hat B_{\alpha\beta})-\eps^{\alpha\beta\rho\m\n\lambda}~\partial_\m b \partial_\rho \bar H_{\n\lambda}\bar H_{\alpha\beta} ~(\delta b)\Big) \ .
\eea 
The claim is that we still have an extra gauge symmetry parameterized by $\phi$
\bea
\delta \hat B^a_{\m\n}&=& \bar H^a_{\m\n} \phi \ ,\\
\delta b &=&\phi \ ,
\eea  where a is the group index. (If we need to introduce other fields, we simply let
$\delta \mbox{(Other~Fields)}=0$.) To check that, we see that the ordinary derivatives part are cancelled as in the abelian PST action (recall \eqref{vv}). The extra term that only exists in the non-Abelian model is 
\bea
\eps^{\alpha\beta\rho\m\n\lambda}\partial_\m b ~f^a_{bc} A^b_\rho \bar H^c_{\n\lambda} \bar H^a_{\alpha\beta} \phi=0 \ ,
\eea
where $f^a_{bc}$ is the structure constant. We see this term ''vanish identically'' since there are three total antisymmetry indices on $\bar H^c_{\n\lambda}$. Notice that the equation of motion of $b(x)$ again does not give an extra constraint since its field equation vanishes using the self-duality condition. Therefore, after fixing $b=x^5$, from \eqref{npst} we obtain
\bea
S&=&{1\over 4} \int d^6x~Tr~\Big ( -{1\over 2} H^{ij5}H_{ij5}-{1\over 3!} H^{ijk}H_{ijk}+{1\over 2} {{\cal H}^{ij5}} {{\cal H}_{ij5}}\Big)\nn\\
&=&-{1\over 4!} T_{M5}T_{M2}^{-2}\int d^6x~Tr~(2 H_{ijk} H^{ijk} +\epsilon^{ijklm} H_{klm} \partial_5 \hat B_{ij}) \ ,
\eea
which indeed reduces to our previous non-covariant multiple M5-branes  (PS-like) action that gives the self-duality condition for non-Abelian 2-form potentials.
\\

However, now we are going to see that the main challenge is to have a covariant version of the non-Abelian 2-form theory (without the self-duality consideration), which means that we will see that it is unclear how to define $H_{\m\n\lam}$ such that $L\sim H_{\m\n\lam}H^{\m\n\lam}$ is Lorentz covariant. 
\\

Naturally, one would introduce another Abelian auxiliary field $ a_\m(x)$ to define the covariant non-Abelian 2-form theory and the previous non-covaraint algebra (gauge transformation laws and 3-form field strengths) could be obtained by an extra gauge fixing (if a new gauge symmetry exists). Let us try this approach. 
\\

First of all, we need to consider that a five dimensional one-form potential $A(x^i)$ appears as the new field replacing $RB^{(o)}_{i5}$ in large R limit as mentioned earlier. We again separately define the zero modes when reducing to multiple D4-branes. Let $\m,\n,\lambda...= 0,1,2,3,4,5$ and $ i,j,k..=0,1,2,3,4$ and we again use the non-local operator which only consistently acts on the KK-modes that dominate contributions in 6D (In the following, we will ignore the index of (KK) notation).  We now define a covariant non-Abelian gauge transformation of 2-form potentials in 6D 
\bea
\delta B_{\m\n}=[D_\m,\Lam_\n]-[D_\n,\Lam_\m]+[B_{\m\n},\lambda]-[F_{\m\n},(a \cdot \del)^{-1}a^\rho \Lam_\rho] \label{KW} \ ,
\eea 
where $a^\mu (x)$ is the U(1) auxiliary spacetime-dependent field that is used to covariantize the gauge algebra. (Notice that $a(x)$ is not $b(x)$. The field $b(x)$ was used to covariantize the non-Abelian self-dual theory.) 
Here the covariant derivative is defined in terms of the one-form potential
\bea
D_\m=\del_\m+A_\m\ .
\eea We then need to search for a new gauge symmetry that allows us to gauge $A_5$ away in order to be consistent with our non-covariant gauge algebra. It could be also interesting to understand more about the physical meaning of this 1-form potential such as what happens when it comes to the compactification so that the role of A is mapped to the zero modes, $B^{(0)}_{i5}$. We notice that if A field is determined by other fields it will  contradict a recent result that the 6D theory is trivial when one only uses (2,0) tensor multiplets \cite{yu}. Thus, it seems one needs to consider this A-field as a new field in the interacting 6D theory. We conjecture that these new degrees of freedom should relate to the condensation of tensionless strings living in M5-branes. We will discuss this A-field more later and in particular we see hints that this A-field is a five-dimensional field hence it does not influence degrees of freedom in the six-dimensional spacetime.
\\

We define $F_{\m\n}=[D_\m,D_\n],~\delta A_\m(x)=[D_\m,\lam],~\delta a_\m(x)=0$.  Any coupling constant should be be introduced due to the conformal nature of M5-branes. To be justified as the deformation of the Abelian 2-form theory, the gauge transformation of B-field should be invariant under the ''redudent transformation'' now defined in a covariant way:
\bea
\delta \Lam_\m=[D_\m,\a] \ .
\eea 
We find that this symmetry exists only if 
\bea
a^\m A_\m=0 \label{a} \label{cc}
\eea 
The consequence of this condition is that if we fix a direction for a, says the 5-th direction, then the component $A_5$ will decouple from the theory, which is consistent with the previous non-covariant model where $A_5$ is absent in the beginning.  Notice that the gauge transformation on the condition \eqref{cc} gives
\bea
a^\m [D_\m.\lambda]= (a\cdot \partial) \lambda =0  \ .
\eea 
As a consistency constraint on $\lam$. This tells us $\lambda$ is a five-dimensional parameter, as expected. One can check that the gauge transformation \eqref{KW} is closed:
\bea
[\delta_1,\delta_2]=\delta_3
\eea with
\bea
\lambda_3&=&[\lambda_1,\lambda_2]\\
\Lambda_{\m3}&=&[\Lambda_{\m1},\lambda_2]-[\Lambda_{\m2},\lambda_1] \ .
\eea It is a positive hint that it is possible to have a covariant version of the non-Abelian 2-form gauge theory. 
\\

We start, however, meeting problems when we try to  define covariant 3-form field strengths. If we adopt the forms suggested by the previous non-covariant  3-form field strengths, we should define
\bea
H_{\m\n\lam}&=&[D_\m, B_{\n\lam}]+[D_\n, B_{\lam\m}]+[D_\lam, B_{\m\n}]+[F_{\m\n},(a\cdot \del)^{-1}a^\rho B_{\lam\rho }]\nn\\
&+&[F_{\n\lam},(a\cdot \del)^{-1}a^\rho B_{\m\rho }]+[F_{\lam\m},(a\cdot \del)^{-1}a^\rho B_{\n\rho}] \ .
\eea
If we want this form of the field strength to gauge-transform covariantly, that is,
\bea
\delta H_{\m\n\lam}=[H_{\m\n\lam},\lam] \ ,
\eea
we find that we first will need 
\bea
a^\m F_{\m\n}&=&0 \label{b}
\eea which is consistent with the fact that $A_\m$ is a five-dimensional field after a preferential choice of the vector $a_\m(x)$.
However, we will also need to impose constraints on gauge parameters $\Lam_\m$ and $\lam$
\bea
(a\cdot \partial)[D_\m,\lambda]&=&0 \ ,\\
(a\cdot \partial)^{-1}a^\rho [D_\lambda, \Lambda_\rho]&=& [D_\lambda, (a \cdot \partial)^{-1}a^\rho \Lambda_\rho] \ .
\eea
We see the problem is that these constraints on gauge parameters are too strong and it is almost saying that the field $a_\m$ is constant, which means that there is no manifest Lorentz symmetry.  Furthermore, we do not want to set equations \eqref{a}, \eqref{b} by hand that contradicts the action principle. The two constraints on fields,
$
a^\m A_\m=0,
a^\m F_{\m\n}=0
$ should be solved simultaneously. One can, for instance, solve them by considering
\bea
a_\lam&=&\Big(\delta^\n_\lam-x^{-2} x_\lam x^\n\Big) B_\n \ , \\
A^a_\lam&=& x_\lam [1-(a\cdot\del)^{-1}(a\cdot\del)]C^a \ ,
\eea where a is the group index and $x_\lam$ is the spacetime coordinate.  We introduce a field $B_\n$ (Abelian field), and also a field $C^a$.  However, these solutions means that the degrees of freedom of A are too small.   Another problem is that we still need to find a ''redundant'' hidden symmetry for $a(x)$ that allows us to reduce:
\bea
\delta B_{\m\n}=[D_\m,\Lam_\n]-[D_\n,\Lam_\m]+[B_{\m\n},\lambda]-[F_{\m\n},(a \cdot \del)^{-1}a^\rho \Lam_\rho] \ ,
\eea to
\bea
\delta B_{ij}=[D_i,\Lam_j]-[D_j,\Lam_i]+[B_{ij},\lambda]-[F_{ij},\del^{-1}_5 \Lam_5] \ .
\eea And this symmetry (if exists) should also allow us to reduce
\bea
H_{\m\n\lam}&=&[D_\m, B_{\n\lam}]+[D_\n, B_{\lam\m}]+[D_\lam, B_{\m\n}]+[F_{\m\n},(a\cdot \del)^{-1}a^\rho B_{\lam\rho }]\nn\\
&+&[F_{\n\lam},(a\cdot \del)^{-1}a^\rho B_{\m\rho }]+[F_{\lam\m},(a\cdot \del)^{-1}a^\rho B_{\n\rho}] \ ,
\eea to
\bea
H_{ijk}&=&[D_i, B_{jk}]+[D_j, B_{ki}]+[D_k, B_{ij}]+[F_{ij},\del_{5}^{-1} B_{k5}]\nn\\
&+&[F_{jk},\del_{5}^{-1} B_{i5}]+[F_{ki},\del_{5}^{-1} B_{j5}] \ .
\eea 
At present we do not know the full answer to these questions. 

\newpage
\subsection{Conclusion}

In this thesis, we study various formulations of chiral fields in different dimensions, including non-covaraint and covariant formulations. We mostly focus on the formulation of chiral two-form in six-dimensions.  After reviewing the Abelian theory, we construct gauge transformation laws, equations of motion and an action for its non-Abelian generalization. The new ingredient as compared to previous approaches is the non-locality in the compact direction. We expect that this model could be used to describe multiple M5-branes  (gauge sector) on $S^1$ with a finite radius R.
\\

We also comment on this model's decompactified limit and compare it with other approaches on M5-branes system.  We apply the BRST-antifield gauge-fixing method on this non-Abelian 2-form gauge theory and generalize the algebra to construct a non-Abelian 3-form gauge theory. Finally we point out difficulties when trying to find a manifest Lorentz-covariant theory by utilizing auxiliary fields.  
\\

The main open problems are
\begin{enumerate}
\item
A well defined $R\rightarrow \infty$ limit of the theory with the 6D Lorentz symmetry.
\item
The supersymmetric extension of the theory.
\item
The quantization of the theory.
\end{enumerate}
It is fair to say that the full structure of M5-branes is still highly mysterious and further investigation is certainly needed.

\newpage
\section*{Acknowledgments}
I would like to express my gratitude to my advisor, Pei-Ming Ho, 
for his guidance and for introducing me to the study of M5-branes theories. I appreciate his inspiration and the great efforts to explain everything clearly. I benefited from many people when discussing this topic:  Wei-Ming Chen, Yen-Ta Huang, Sheng-Lan Ko, Fech-Scen Khoo, Yi-Chuan Lu, Chen-Te Ma, Yutaka Matsuo, Chi-Hsien Yeh and Chen-Pin Yeh.  I am also grateful to Chuan-Tsung Chan and  Wen-Yu Wen for being my thesis defense committee. Finally, I would like to thank my parents, Wung-Hong Huang and Li-Hua Lai, for their constant support and encouragement.

\end{CJK} 
\end{document}